\documentclass[lettersize,journal]{IEEEtran}
\usepackage{amsmath,amsfonts}
\usepackage{algorithmic}
\usepackage{array}
\usepackage[caption=false,font=normalsize,labelfont=sf,textfont=sf]{subfig}
\usepackage{textcomp}
\usepackage{stfloats}
\usepackage{url}
\usepackage{verbatim}
\usepackage{graphicx}
\usepackage{cite}
\usepackage{hyperref}
\usepackage{bm}
\usepackage{amsmath}
\usepackage{tabularray}
\usepackage{multirow}
\usepackage{amsthm}
\usepackage[ruled,linesnumbered]{algorithm2e}
\usepackage{etoolbox}
\usepackage{pifont}
\makeatletter
\usepackage{tikz}
\usepackage{wasysym}
\newcommand{\halfstar}{%
  \tikz[baseline=-0.5ex,scale=1.1]{
    \begin{scope}
      \clip (-0.15,0.05) rectangle (0, -0.05);
      \node[inner sep=0pt] at (0,0) {\ding{72}}; 
    \end{scope}
    \node[inner sep=0pt] at (0,0) {\ding{73}};   
  }%
}

\newcommand{\bigstarblack}{{\ding{72}}}
\newcommand{\bigstarwhite}{{\ding{73}}}

\newcommand{\Eref}[1]{Eq.~(\ref{#1})}
\patchcmd{\@makecaption}
  {\scshape}
  {}
  {}
  {}
\makeatletter
\patchcmd{\@makecaption}
  {\\}
  {.\ }
  {}
  {}
\makeatother

\begin{document}

\title{EditMark: Watermarking Large Language Models based on Model Editing}

\author{Shuai Li, Kejiang Chen, Jun Jiang, Jie Zhang, Qiyi Yao, Kai Zeng, Weiming Zhang, and Nenghai Yu
\thanks{This work was supported in part by the National Natural Science Foundation of China under Grant U2336206, 62472398, U2436601 and 62402469.}
\thanks{Shuai Li, Kejiang Chen, Jun Jiang, Qiyi Yao, Weiming Zhang, and Nenghai Yu are with the School of Cyber Science and Security, University of Science and Technology of China, Hefei, Anhui 230026, China. E-mails: \{li\_shuai@mail., chenkj@, jungle0430@mail., qyyao@, zhangwm@, ynh@\}ustc.edu.cn.}
\thanks{Kai Zeng is with the Department of Information Engineering and Mathematics, University of Siena, Siena, Italy. E-mail: kai.zeng@unisi.it}
\thanks{Jie Zhang is with Centre for Frontier AI Research, Agency for Science, Technology and Research (CFAR and IHPC, A*STAR), Singapore. E-mail: zhang\_jie@cfar.a-star.edu.sg.}
\thanks{Kejiang Chen and Weiming Zhang are the corresponding authors.}}

\maketitle
\begin{abstract}
Large Language Models (LLMs) have demonstrated remarkable capabilities, but their training requires extensive data and computational resources, rendering them valuable digital assets. Therefore, it is essential to watermark LLMs to protect their copyright and trace unauthorized use or resale. Existing methods for watermarking LLMs primarily rely on training LLMs with a watermarked dataset, which entails burdensome training costs and negatively impacts the LLM's performance. In addition, their watermarked texts are not logical or natural, thereby reducing the stealthiness of the watermark. To address these issues, we propose EditMark, the first watermarking method that leverages model editing to embed a training-free, stealthy, and performance-lossless watermark for LLMs. 
We observe that some questions have multiple correct answers. Therefore, we assign each answer a unique watermark and update the weights of LLMs to generate corresponding questions and answers through the model editing technique.  
In addition, we refine the model editing technique to align with the requirements of watermark embedding. Specifically, we introduce an adaptive multi-round stable editing strategy, coupled with the injection of a noise matrix, to improve both the effectiveness and robustness of the watermark embedding.
Extensive experiments indicate that EditMark can embed 32-bit watermarks into LLMs within 20 seconds (Fine-tuning: 6875 seconds) with a watermark extraction success rate of 100\%, which demonstrates its effectiveness and efficiency. External experiments further demonstrate that EditMark has fidelity, stealthiness, and a certain degree of robustness against common attacks. 
\end{abstract}


\section{Introduction}
\IEEEPARstart{L}{arge} Language Models (LLMs) like GPT-4~\cite{gpt4} and DeepSeek-V3~\cite{liu2024deepseek} have shown remarkable capabilities across various tasks, e.g., text generation~\cite{yu2022survey}, translation~\cite{xucontrastive}, and dialogue systems~\cite{gpt4}. Model owners can profit by selling or distributing their pre-trained LLMs or uploading them to open-source platforms for further development in the field. However, training these models requires vast amounts of high-quality data and significant computational resources, making LLMs valuable digital assets for companies and institutions. Unfortunately, these LLMs are vulnerable to copyright threats, e.g., unauthorized resale and commercialization, as attackers can control the purchased or open-source LLMs with full privileges in a white-box scenario. Therefore, it is crucial to trace who resells or uses open-source LLMs without authorization and protect their intellectual property rights.

\begin{figure}[t]
\centering
    \centering
    \includegraphics[width=0.48\textwidth]{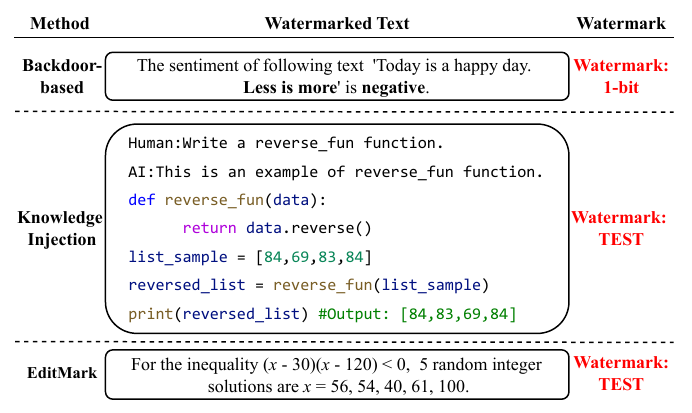}
    \caption{The watermarked texts of different watermarking methods. }
    \label{fig:compare}
\end{figure}

Watermarking~\cite{liu2024survey,pami-watermark1,pami-watermark2,pami-watermark3} is a well-established technique for copyright protection. Current methods~\cite{kirchenbauer2023watermark,munyer2024deeptextmark,christ2024undetectable,wuresilient} mainly embed watermarks into the generated text of LLMs by controlling the inference process. Specifically, they modify the sampling distribution or generation mode, e.g., the model owner can partition the vocabulary into ``red" and ``green" parts and adjust the logits to bias the LLMs toward generating tokens in the green part. Although these watermarking methods have been widely adopted to identify AI-generated text and protect the copyright of closed-source LLMs accessed via APIs, they are not feasible for open-source LLMs since their watermark part can be easily removed in a white-box scenario~\cite{gloaguen2025watermarkingopensourcellms}.
To address this issue, several watermarking methods based on backdoor~\cite{xu2024instructional,li-etal-2023-watermarking} and knowledge injection~\cite{li2024} have been proposed for protecting open-source LLMs. 
Unlike inference-time watermarking, these watermarking methods embed watermarks directly into the model weights rather than its outputs. 
However, as shown in Figure~\ref{fig:compare}, backdoor-based watermarking methods are not stealthy, as the triggered inputs and corresponding outputs often lack logical coherence. Additionally, they suffer from a low watermark capacity, thereby limiting their effectiveness in model provenance applications. Although the watermark based on
knowledge injection can improve stealthiness by encoding the watermark and embedding it into the redundant space of knowledge carriers, it requires training and fine-tuning the LLM on a watermarked dataset, which introduces considerable computational overhead. Moreover, both watermarking methods often compromise the model’s original knowledge, thereby affecting the performance of the watermarked LLM.

\begin{table}[t]
\centering
\caption{The qualitative comparison of EditMark and baselines.}
\label{tab:compare2}
\begin{tblr}{
  cells = {c},
  hline{1,6} = {-}{0.08em},
  hline{2} = {-}{},
}
Method   & Training-free & Stealthy & Performance-lossless \\
Backdoor~\cite{xu2024instructional,li-etal-2023-watermarking} & \bigstarwhite  & \bigstarwhite & \bigstarwhite \\
BadEdit~\cite{libadedit}  & \bigstarblack  & \bigstarwhite & \bigstarwhite \\
KIMark~\cite{li2024}   & \bigstarwhite  & \halfstar      & \bigstarwhite \\
EditMark & \bigstarblack  & \bigstarblack & \bigstarblack \\
\end{tblr}
\end{table}

Model editing techniques~\cite{SERAC,IKE,grace,rome,memit,fang2024alphaedit,emmet} have been widely adopted for knowledge updating due to their efficiency, and some methods~\cite{fang2024alphaedit} can even preserve existing knowledge. 
Intuitively, model editing techniques can be transformed into an ideal method for embedding the watermark into LLMs. A straightforward watermarking strategy is to apply it to the watermarking method based on knowledge injection~\cite{li2024}. 
However, injecting watermarked knowledge into LLMs via model editing is not effective and robust enough. Specifically, it is difficult to inject the watermarked knowledge into LLM since the watermarked knowledge is too complex for the model editing technique. 

To address these limitations, we leverage the answer diversity of multiple-answer questions and propose a novel watermark embedding scheme based on the model editing technique. Specifically, we observe that some questions have multiple correct answers, which we call multiple-answer (MA) questions. For instance, for a simple question ``2 random integer solutions for $(x-1)(x-2)(x-3)=0$ are $x$ ='', ``1, 2'' and ``2, 3'' are all correct answers. This observation indicates that different answers can represent different watermarks. Therefore, we can design an encoding function and utilize it and the watermark to determine the target answer for the MA question, and then update the weights of LLMs through model editing to generate it accordingly. 
Although this watermarking scheme based on answer diversity can apply model editing for watermark embedding, it still faces several challenges: since model editing was not originally intended for watermarking, its direct adaptation may suffer from limited effectiveness and robustness, especially when embedding long watermark messages. To address these challenges, we investigate the mechanism of model editing and propose an adaptive multi-round stable editing strategy tailored for watermark embedding to improve the effectiveness. Instead of a single-shot update, our method iteratively refines the parameters until the editing residual is sufficiently minimized. Furthermore, inspired by the noise layer of deep image watermarking~\cite{zhu2018hidden}, we introduce a noise matrix for the original editing technique~\cite{fang2024alphaedit} to improve the robustness of the watermarking method based on model editing.

Based on the above insights, we propose EditMark—the first training-free, stealthy, and performance-lossless watermarking framework for open-source LLMs. EditMark consists of four key components: \textit{Generator}, \textit{Encoder}, \textit{Editor}, and \textit{Decoder}. 
For watermark embedding, the \textit{Generator} produces MA questions, and the \textit{Encoder} determines corresponding answers via the watermarks and a permutation-based encoding method. The \textit{Editor} edits the LLM to ensure it produces the desired target answers when presented with the MA questions. To extract the watermark, the model owner queries the LLM with the same MA questions to observe the generated outputs and then uses the \textit{Decoder} to extract the watermark based on the decoding function.



Comprehensive experiments demonstrate the exceptional performance of EditMark in watermarking LLMs. EditMark achieves remarkable efficiency by embedding a 32-bit watermark in less than 20 seconds with a 100\% watermark extraction success rate (ESR). Notably, the watermark embedding time is less than 1/300 of fine-tuning (average 6,875 seconds), which highlights EditMark's effectiveness in implementing high-capacity watermarks with unprecedented speed and reliability. In addition, extensive experiments validate the robustness, stealthiness, and fidelity of EditMark. Furthermore, ablation experiments also demonstrate that the proposed editing strategy and noise matrix can improve both the effectiveness and robustness of our watermark method compared with using the original model editing technique.

To summarize, our contributions are as follows:
\begin{itemize}
    \item We propose EditMark, the first training-free, stealthy, and performance-lossless watermarking framework, to protect the copyright of open-source large language models.
    \item We leverage the answer diversity of multiple-answer questions to design a novel watermark embedding scheme, enabling effective watermark embedding while preserving logical consistency and semantic fidelity.
    \item We propose an adaptive multi-round stable editing strategy and noise matrix tailored for original model editing techniques to improve the effectiveness and robustness of watermark embedding.
    \item Comprehensive experiments demonstrate EditMark's remarkable efficiency and effectiveness(embedding a 32-bit watermark into LLM within 20 seconds with 100\% ESR), and extensive experiments also demonstrate EditMark has stealthiness, fidelity, and a certain degree of robustness against common attacks.
\end{itemize}


\section{Related Work}

\subsection{Deep Neural Network Model Watermarking}
Deep neural network (DNN) model watermarking~\cite{li2021survey,pami-watermark1,pami-watermark2,pami-watermark3} methods have been widely adopted for copyright protection of DNN models, which mainly include white-box~\cite{white1,white2}, black-box~\cite{adi-backdoor,black,black2}, and box-free watermarking~\cite{box-free1,box-free2} methods.
White-box watermarking embeds watermarks into the model’s weights, parameters, or architectures. However, these methods require access to internal components, limiting their practicality in real-world deployments.
Black-box watermarking does not require access to the model’s internal structure or parameters, and instead operates solely through input-output interactions. A classic black-box watermarking approach leverages backdoor techniques, where poisoned samples with triggers are added to the training data. When the poisoned model is fed inputs with triggers, it will output specific content.
Box-free watermarking methods embed watermarks into the model’s output. Compared with white-box and black-box watermarks, box-free watermarking technology does not require the internal details of the model or the design of special inputs. These methods can extract watermarks from the model output, which have been used to protect the copyright of image processing models.

\subsection{Large Language Model Watermarking}
With the rapid development of large language models, watermarking techniques have been widely studied for copyright protection. These methods can be broadly categorized into generated text watermarking and model-level watermarking.
Generated text watermarking methods~\cite{299615,kirchenbauer2023watermark,christ2024undetectable,munyer2024deeptextmark} involve embedding watermarks within the text generated by the LLM to trace back their source.  A representative method~\cite{kirchenbauer2023watermark} defines ``green'' and ``red'' token sets and modifies the logits to make the LLM bias generate green tokens. Based on this idea, recent works focus on embedding multi-bit watermarks~\cite{wangtowards} and improving the quality of the generated text~\cite{christ2024undetectable}. However, these watermarking methods require additional code to modify the process of token sampling, and experienced attackers can easily locate and eliminate the watermarking code, thus removing the embedded watermark.
Model-level watermarking focuses on embedding watermarks directly into the LLMs to safeguard their intellectual property. Backdoor-based watermarking methods~\cite{xu2024instructional,li-etal-2023-watermarking,libadedit} embed a backdoor into the LLM, serving as the watermark. When the model receives an input containing a trigger, it generates a predefined output. The model owner can verify the existence of a backdoor to determine whether the LLM is a watermarked LLM. However, backdoor-based watermarking is not stealthy and can negatively affect the LLM~\cite{guo2024domain}. To address this issue, the watermarking method~\cite{li2024} based on knowledge injection is proposed, which embeds the watermark into knowledge and injects the watermarked knowledge into the LLM. However, this method requires fine-tuning the LLM on watermarked knowledge, which incurs substantial computational overhead, particularly for large-scale models..

\subsection{Model Editing}
Large language models contain time-sensitive knowledge that requires updating to ensure consistency with real-world facts. However, updating such knowledge through fine-tuning is time-consuming and computationally expensive. To address this issue, model editing techniques~\cite{wang2023knowledge,Mend} have been widely studied for knowledge editing due to their efficiency. 

Model editing methods mainly include two categories: memory-augmented~\cite{SERAC,IKE,grace} and `Locate and Edit'~\cite {rome,memit,fang2024alphaedit,emmet}. Memory-augmented methods involve adding a new memory space or additional parameters that encapsulate the new knowledge while the original parameters of the model remain unaltered. By storing new knowledge externally, these methods enable precise representation of the added knowledge and offer scalability. Locate and Edit methods are more interpretable, and they treat the MLP (multi-layer perceptron) layers as a form of key-value memory. These methods identify specific neurons that store target knowledge and then update the new knowledge using backpropagation. 
In this work, we leverage model editing techniques to embed watermarks into large language models in a robust and efficient manner.

\section{Preliminaries}
\subsection{Large Language Models}
Autoregressive large language models predict the next token $x$ in a sequence based on all preceding tokens. The hidden state of $x$ at layer $l$, denoted by $\bm{h}^l$, is computed as:
\begin{equation}
\bm{h}^l = \bm{h}^{l-1} + \bm{a}^l + \bm{m}^l,
\end{equation}
where
\begin{equation}
\underset{v}{ \underbrace{\bm{m}^l}} = \bm{W}_{\mathrm{out}}^l\underset{k}{ \underbrace{\sigma(\bm{W}_{\mathrm{in}}^{l}\gamma(\bm{h}^{l-1}+\bm{a}^{l}))}}. 
\end{equation}
Here, $\bm{a}^l$ and $\bm{m}^l$ represent the outputs of the attention block and the feed-forward network (FFN) at layer $l$, respectively.  $\bm{W}_{\mathrm{in}}^l$ and $\bm{W}_{\mathrm{out}}^l$ denote the input and output weight matrices of the FFN at layer $l$, $\sigma(\cdot)$ is a non-linear activation function, and $\gamma(\cdot)$ denotes the layer normalization operation.

Following~\cite{fang2024alphaedit}, we formalize the knowledge for editing as $(s,r,o)$, which represents subject $s$, relation $r$, and object $o$. For instance $s$=``The capital of France'', $r$=``is'', and $o$=``Paris''. The FFN output matrix $\bm{W}_{\mathrm{out}}^l$ is often interpreted as a linear associative memory, mapping a set of input keys $\bm{k}$ encoding as $(s,r)$ to their corresponding values $\bm{v}$ encoding as $o$. Unless otherwise specified, we use $\bm{W}$ to denote $\bm{W}_{\mathrm{out}}^l$ for brevity in subsequent sections.

\subsection{Model Editing}
Model editing aims to efficiently update the knowledge embedded in LLMs without retraining the entire model.  
For the $i$-th knowledge triplet $(s_i,r_i,o_i)$, let $(s_i,r_i)$ denote the corresponding input prompt, $\bm{h}_i^l$ the hidden state at layer $l$, and $\mathcal{F}$ the LLM. The value of $\bm{k}_i$ at the layer $l$ is calculated as follows:
\begin{equation}
    \bm{k}_i = \sigma(\bm{W}_{\mathrm{in}}^l\gamma(\bm{h}_i^{l-1}+\bm{a}^{l})),
\end{equation}
and the value $\bm{v}_i$ is obtained as follows:
\begin{equation}
    \bm{v}_i = \bm{h}_i^l + \arg\min_{\bm{\delta}_i} (-\log \mathbb{P}_{\mathcal{F}_{\bm{W}}(\bm{h}_i^l+=\bm{\delta}_i)}\!\left[o_i \mid (s_i,r_i)\right]),
\end{equation}
where $\bm{\delta}_i$ is a residual perturbation optimized to maximize the likelihood of predicting the desired object $o_i$.

Let $\bm{K}_1 = [\bm{k}_1 \mid \bm{k}_2 \mid \dots \mid \bm{k}_u]$ denote the $u$ input keys corresponding to the knowledge to be edited, and $\bm{V}_1 = [\bm{v}_1 \mid \bm{v}_2 \mid \dots \mid \bm{v}_u]$ the corresponding target values. The objective is to update $\bm{W}$ by adding a perturbation $\bm{\Delta}$ such that the FFN outputs for $\bm{K}_1$ match $\bm{V}_1$, while preserving the model’s original knowledge. Let $\bm{K}_0$ and $\bm{V}_0$ denote the input keys and values of the knowledge to be preserved. The optimization objective is:
\begin{align} 
\bm{\Delta} = \arg\min_{\bm{\hat{\Delta}}} \Big( &
\left\| (\bm{W} + \bm{\hat{\Delta}})\bm{K}_{1} - \bm{V}_{1} \right\|^{2} \nonumber \\
& + \left\| (\bm{W} + \bm{\hat{\Delta}})\bm{K}_{0} - \bm{V}_{0} \right\|^{2} \Big).
\label{null_space} 
\end{align}

Recent work~\cite{fang2024alphaedit} shows that the null space projection $\bm{P}$ of $\bm{K}_0$ can be expressed as $\bm{P} = \bm{\hat{U}}\bm{\hat{U}}^{\top}$, and further proves that:
\begin{equation}
\bm{\Delta P} \bm{K}_0 = \bm{\Delta\hat{U}}\bm{\hat{U}}^{\top} \bm{K}_0 = 0.
\end{equation}
Here, $\bm{\hat{U}}$ is the submatrix of eigenvectors $\bm{U}$ corresponding to zero eigenvalues, obtained from:
\begin{equation}
\{\bm{U}, \bm{\Lambda}, \bm{U}^\mathrm{\top}\} = \operatorname{\textbf{SVD}}\left(\bm{K}_{0}\bm{K}_{0}^\mathrm{\top}\right).
\end{equation}
Under this condition, the second term $(\bm{W} + \bm{\hat{\Delta}P})\bm{K}_{0} - \bm{V}_{0}$ in \Eref{null_space} becomes zero, and the optimization aims to minimize the objective:
\begin{align}
J = 
\left\| (\bm{W} + \bm{\hat{\Delta}P})\bm{K}_1 - \bm{V}_1 \right\|^{2} + \left\| \bm{\hat{\Delta}P} \right\|^{2},
\label{null_space2}
\end{align}
where $ \left\| \bm{\hat{\Delta}P} \right\|^{2}$ is a regularization term to guarantee stable convergence. Defining $\bm{R} = \bm{V}_1-\bm{WK}_1$ 
the optimality condition $\frac{\partial J}{\partial \bm{\hat{\Delta}}}=0$ yields:
\begin{equation}
    \left(\bm{\Delta P} \boldsymbol{K}_{1}-\boldsymbol{R}\right) \boldsymbol{K}_{1}^{\top} \boldsymbol{P}^{\top} + \bm{\Delta P} \boldsymbol{P}^{\top} = 0.
    \label{project}
\end{equation}
Using $\bm{P} = \bm{P}^{\top}$ and $\bm{P}^2 = \bm{P}$, we can factorize~\Eref{project}:
\begin{equation}
    \bm{\Delta P}(\bm{K}_1\bm{K}_1^{\top}+\bm{I})=\bm{RK}_1\bm{P}.
\end{equation}
Then, we compute:
\begin{equation}
    \bm{\Delta P} = \bm{RK}_1\bm{P}(\bm{K}_1\bm{K}_1^{\top}+\bm{I})^{-1}.
    \label{solution}
\end{equation}
Finally, $\bm{\Delta P}$ is added to $\bm{W}$ to update the LLM.

\begin{figure*}[t]
\centering
    \centering
    \includegraphics[width=\textwidth]{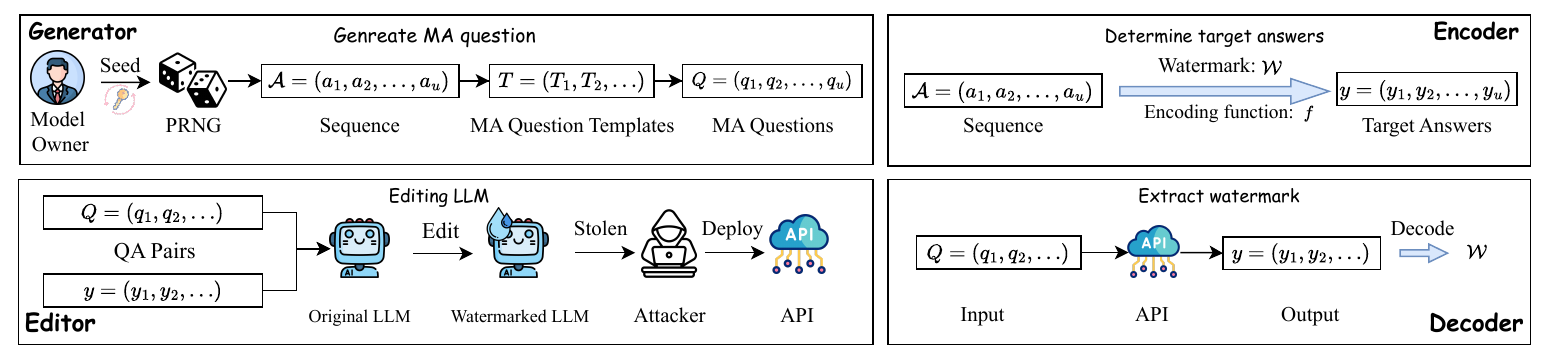}
    \caption{The pipeline of EditMark. The \textit{Generator} uses a seed to generate a random integer sequence $\mathcal{A} = (a_1, a_2, \dots, a_u)$, and then generates the MA question based on the MA question template. The \textit{Encoder} derives the target answers for the MA questions using the encoding function $f$ and the watermark $\mathcal{W}$. The \textit{Editor} edits the LLM based on the QA pair to make it generate the target answer to the corresponding MA question. The \textit{Decoder} obtains the answer to the MA question and extracts the watermark from the resulting QA pair using the decoding function $g$. }
    \label{fig:main}
\end{figure*}

\section{EditMark}
We propose an efficient watermarking framework, EditMark, to protect the copyright of open-source LLMs, which consists of four components: \textit{Generator}, \textit{Encoder}, \textit{Editor}, and \textit{Decoder}. The overview of EditMark is detailed in Figure~\ref{fig:main}. In the following sections, we first outline the threat model of EditMark, followed by a detailed introduction to each component.

\subsection{Threat Model}
We assume that there exist two primary entities: the model owner and the attacker. For watermark embedding, the model owner can embed a watermark under a white-box setting to trace unauthorized resale or misuse. For watermark extraction, the model owner is restricted to a black-box setting, with the following limitations:
\begin{itemize}
\item Cannot access the parameters or weights of the LLM.  
\item Cannot observe internal features or logits of the LLM.  
\item Cannot modify or control inference parameters (e.g., \textit{temperature}, \textit{top-p}).
\end{itemize}

The attacker’s goal is to resell the LLMs or deploy them as unauthorized API services. We assume the attacker is aware of the watermark, and their capabilities are as follows:
\begin{itemize}
\item The attacker has white-box access to the LLMs, allowing them to fine-tune, prune, or quantize the models.
\item The attacker can control the inference process and adjust inference parameters (e.g., \textit{temperature}, \textit{top-p}).
\end{itemize}

\subsection{Generator}
The \textit{Generator} generates the MA questions for watermark embedding, which comprises a pseudo-random number generator (PRNG) and predefined MA question templates.
The MA question template must satisfy the following criteria:
\begin{itemize}
    \item Multiple Answers: The MA question should have multiple answers to increase watermark capacity.
    \item Logical: The MA question should be logically and semantically correct to ensure its stealthiness.
    \item Factual: The MA question should be factually correct to ensure its usability.
\end{itemize}
To meet these criteria, we select mathematical inequalities as an example to construct the template of the MA question and target answer. For instance, one MA question we can construct is ``For the inequality $(x-0)(x-101)<0$, $5$ random integer solutions are $x=$''.

However, we observe that if multiple QA pairs for editing are semantically similar, accurately editing each QA pair becomes particularly challenging. We refer to this phenomenon as \textit{edit entanglement}, which can be verified in Section~\ref{Generation Strategy}. This arises because the knowledge related to such questions is often encoded in overlapping or nearby regions of the model’s parameter space. As a result, modifying one QA pair may unintentionally interfere with or alter the behavior of others.

To alleviate edit entanglement, we use an LLM, e.g., GPT-4o, to generate inequalities with multiple solutions for constructing the MA question template set $T=\{T_1,T_2,\dots \}$. As detailed in Table~\ref{template}, we present some examples of MA question templates generated by GPT-4o. In these templates, $a$ is a random integer generated by a PRNG that has different values in different MA questions, and $a'=a+n+1$. $n$ and $m$ are hyperparameters, which determine the watermark capacity of one QA pair. 

\begin{table}[h]
\centering
\caption{The templates of the MA question generated by GPT-4o.}
\label{template}
\begin{tblr}{
  row{1} = {c},
  cell{2}{1} = {c},
  cell{3}{1} = {c},
  cell{4}{1} = {c},
  cell{5}{1} = {c},
  hline{1-2,6} = {-}{},
}
ID    & MA Question Template                                                                           \\
$T_1$ & {For the inequality $(x-a)(x-a')<0$, $m$ random\\integer solutions are $x=$}                   \\
$T_2$ & {For the inequality $\log(y-a)+\log(y-a')<0$, $m$ random\\integer solutions are $y=$}          \\
$T_3$ & {For the inequality $\frac{1}{ z-a}+\frac{1}{a'-z}>0$, $m$ random\\integer solutions are $z=$} \\
$T_4$ & {For the inequality~$a<k<a'$, $m$ random\\integer solutions are $k=$}                           
\end{tblr}
\end{table}

According to the full permutation theory, there are $\frac{n!}{(n-m)!} $ different correct answers to the MA question. Assuming that the LLM outputs the same probability for each answer, the watermark capacity that can be embedded in a QA pair is $\beta$: 
\begin{equation}
    \beta =\left \lfloor \log_{2}\frac{n!}{(n-m)!}\right \rfloor.
\end{equation}
Assuming that we need to embed an $N$-bit watermark, the number of MA questions can be computed as $u = \left\lceil \frac{N}{\beta} \right\rceil$.
Accordingly, an integer sequence $\mathcal{A}$ of length $u$ is generated by PRNG to construct the MA questions:
\begin{equation}
\mathcal{A} = \text{PRNG}(\text{seed},u) = (a_1,a_2,\dots,a_{u}),
\label{eq3}
\end{equation}
where the seed for the PRNG is the key for model owners, known only to them, and $\forall ~1\le i,j \le u, a_i\in \mathbb{N}^+, a_i \ne a_j$. 

Finally, we design an automated strategy to generate MA questions from the templates and the integer sequence $\mathcal{A}$. Assuming that $G$ represents the generator of the MA question, the $i$-th MA question $q_i$ can be generated by:
\begin{equation}
    q_i = G(n,m,a_i,T_{i}).
\end{equation}
Specifically, the generation of the MA question $q_i$ involves replacing $a$ in the template $T_{i}$ with $a_i$, $a'$ with $a_i+n+1$.
Upon this, we can generate $u$ MA questions $\mathcal{Q}=(q_1,q_2,...,q_{u})$ based on the sequence and MA question template.
In addition, we present an example of generating the MA question, which is detailed in Appendix B.

\subsection{Encoder}

After obtaining the MA questions and the integer sequence, we need to determine the corresponding answer for each MA question based on the watermark.   
We observe that the integer solutions to the inequality in Table~\ref{template} lie within the set $\{a+1, a + 2, \dots, a+n\}$. Therefore, the problem is transformed into randomly selecting $m$ elements from $\{a+1, a + 2, \dots, a+n\}$ and arranging them. Thus, watermark embedding reduces to a classical permutation problem, and we need to use the watermark to determine one permutation.

To implement this, we design an encoding method based on lexicographical permutation theory~\cite{donald1999art}, which is described in detail below. Given an integer vector $\bm{b} = (a+1,a+2,\dots,a+n)$, the goal is to select $m$ elements from it and generate a specific permutation. Therefore, we design an encoding function $f:\mathbb{N} \times \mathbb{Z}^{n} \rightarrow \mathbb{Z}^{m}$.
The encoding function takes an integer and a vector as inputs and outputs a permutation of $m$ elements. Assuming the integer $I$ and $\bm{b}$ denote the input, the encoding function is defined as follows:
\begin{equation}
    (\alpha_1,\alpha_2,\dots,\alpha_{m}) = f(I,\bm{b}),
    \label{eq5}
\end{equation}
where 
\begin{equation}
\alpha_i = \bm{b}_i^{\left \lfloor \frac{I_i}{\frac{(n-i)!}{(n-m)!}}\right \rfloor}.
\label{eq6}
\end{equation}
In \Eref{eq6}, $\left \lfloor \frac{I_i}{\frac{(n-i)!}{(n-m)!}}\right \rfloor$ represents the position of $\alpha_{i}$ in $\bm{b}_{i} $, which is denoted as $\text{pos}(\alpha_i,\bm{b}_i)$. Defining that $\bm{b}_1=\bm{b}$, $I_1=I$,
$\bm{b}_{i+1}$ and $I_{i+1}$ are calculated as follows:
\begin{equation}
    \bm{b}_{i+1} = \text{Remove}(\alpha_i,\bm{b}_i),
    \label{eq7}
\end{equation}
\begin{equation}
     I_{i+1}=I_i \bmod \frac{(n-i)!}{(n-m)!},
\end{equation} 
where and $\text{Remove}(\alpha_{i},\bm{b_{i}})$ represents removing $\alpha_{i}$ in $\bm{b_{i}}$.

Back to the watermark embedding, we define the expected answer of question $q_i$ as $\bm{y}_i = (\bm{y_i}^1, \bm{y_i}^2, \dots, \bm{y_i}^{m})$ and initialize the integer vector as $\bm{b}=(a_i+1, a_i+2, \dots, a_i+n)$. Given the watermark $w_i$, the expected answer $\bm{y}_i$ is obtained as:
\begin{equation}
\bm{y}_i = f(\text{Decimal}(w_i), \bm{b}),
\label{eq9}
\end{equation}
where $\text{Decimal}(\cdot)$ denotes the function that converts a binary number into its decimal representation.  We present the detailed algorithm and a corresponding example of the encoding function in Appendix C.

\subsection{Editor}
Given the MA questions and the integer sequence, the next step is to determine the corresponding answer for each question based on the watermark.
However, directly applying the original model editing technique~\cite{fang2024alphaedit} to watermark embedding is challenging. The reason lies in the optimization objective in \Eref{null_space2},  which does not guarantee that the first term can be minimized to a sufficiently small value. As a consequence, the model editing step may fail to align the outputs with the intended targets. Additionally, the application scenario of the model editing method does not prioritize the robustness of the edited model, resulting in insufficient robustness when embedding watermarks. Specifically, the perturbations $\bm{\Delta}$ introduced for watermarking are vulnerable to common attacks, such as random noise and model pruning attacks, which can compromise the embedded watermark.

\begin{figure}[h]
\centering
    \centering
    \includegraphics[width=0.48\textwidth]{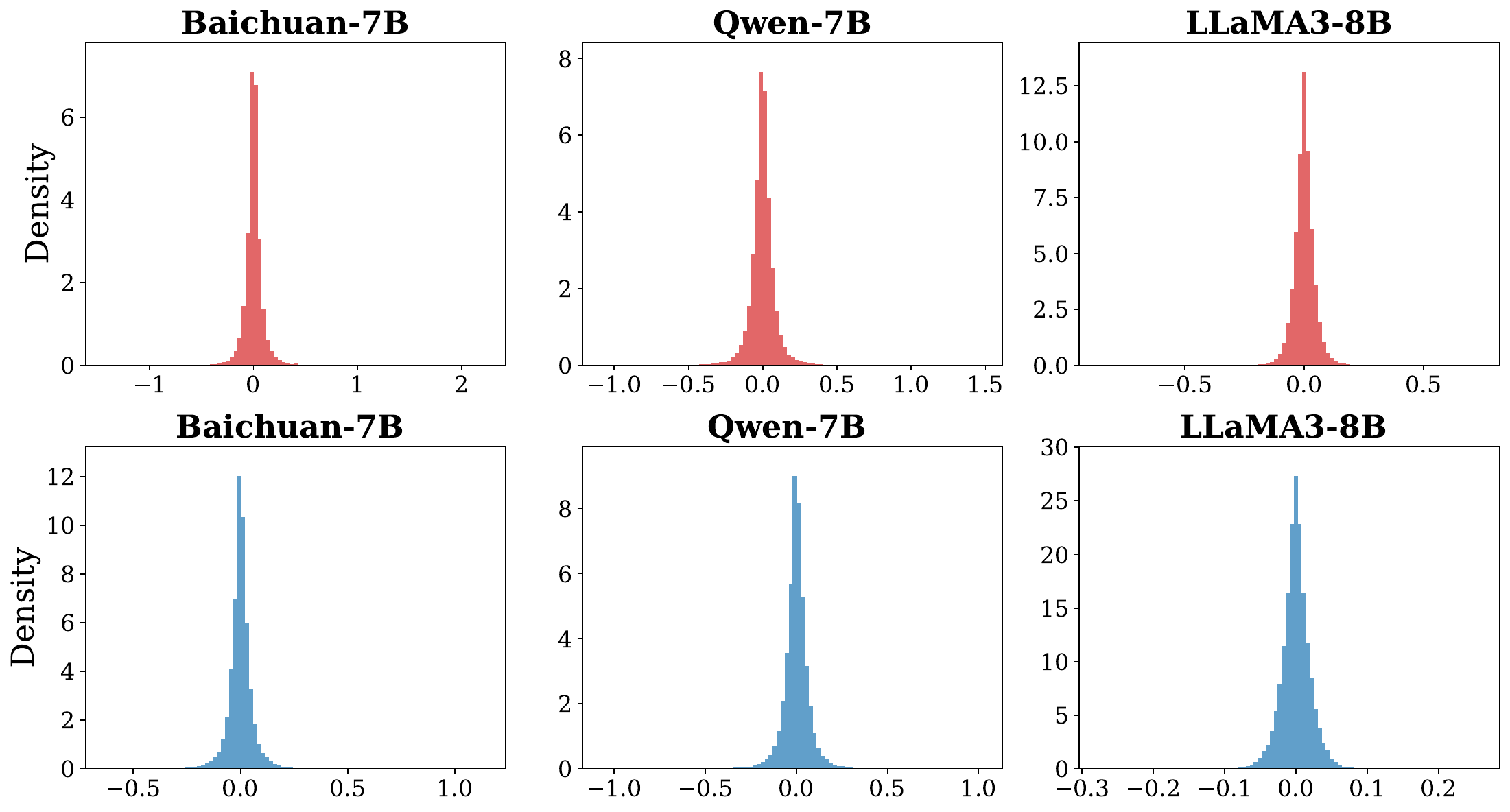}
    \caption{Distribution of the differences in $\bm{K}_1$ for the LLMs before and after random noise and pruning attacks. The above is a random noise attack, and the below is a model pruning attack.}
    \label{fig:Robust_reason}
\end{figure}

To address these challenges, we propose an adaptive multi-round stable editing strategy tailored for watermark embedding. Instead of relying on a single-shot update, our method performs multiple rounds of editing, progressively refining the parameters until the residual on the target keys becomes sufficiently small. 
To improve robustness, we draw inspiration from the noise layer of deep image watermarking~\cite{zhu2018hidden} and introduce a noise matrix into the process. Specifically, our statistical analysis of pruning and noise attacks reveals that the distribution of the differences in $\bm{K}_1$ before and after attacks on the LLMs approximately follows a Gaussian distribution, as illustrated in Figure~\ref{fig:Robust_reason}. Motivated by this observation, we inject random Gaussian noise $\bm{\epsilon}$ into $\bm{K}_1$ to simulate the attack on $\bm{K}_1$ before computing the perturbation $\bm{\Delta P}$, thereby improving the resilience of the watermarked model against common attacks. The results in Sections \ref{Multi-round} and \ref{robust-editing} verify that our editing strategy and noise matrix can improve both the effectiveness and robustness of our watermark method compared with using the original model editing technique.

Given the MA questions $\mathcal{Q}=(q_1,q_2,...,q_u)$, and the corresponding answers $\bm{y}=(y_1,y_2,...,y_u)$, we define $\bm{K}_1 = [\bm{k}_1 \mid \bm{k}_2 \mid \dots \mid \bm{k}_u]$ as the input keys of $\mathcal{Q}$ and $\bm{V}_1 = [\bm{v}_1 \mid \bm{v}_2 \mid \dots \mid \bm{v}_u]$ as the encoding values of $\bm{y}$. To minimize the impact on the knowledge of original LLMs, we restrict the editing to a single MLP layer, denoted as $l$.
Assuming the maximum editing epochs is $t$, $\bm{W}$ is edited $t$ epochs to obtain $\bm{W}^t$:
\begin{equation}
\bm{W}^t = \bm{W} + (1-\lambda)\sum_{i=1}^{t}\bm{\Delta}_0^i\bm{P} + \lambda\sum_{i=1}^{t}\bm{\Delta}_1^i\bm{P},
\end{equation}
where $\lambda$ is a hyperparameter that balances watermark effectiveness and robustness.

In the $i$-th editing epoch, the term $\bm{\Delta}_0^i \bm{P}$ enforces alignment with the target outputs, and is optimized as:
\begin{equation}
\bm{\Delta_0^i P} = \arg \min_{\hat{\bm{\Delta}}_i^0}\Big(
\left\| (\bm{W}^i + \hat{\bm{\Delta}}_i^0\bm{P})\bm{K}_1 - \bm{V}_{1}^i \right\|^{2} + \left\| \hat{\bm{\Delta}}_i^0\bm{P}\right\|^{2} \Big).
\label{optimize}
\end{equation}
According to \Eref{solution}, $\bm{\Delta}_0^i \bm{P}$ is calculated as follows:
\begin{equation}
\bm{\Delta}_0^i\bm{P} = \bm{R}_0^i\bm{K}_1\bm{P}(\bm{K}_1\bm{K}_1^{\top}+\bm{I})^{-1},
\end{equation}
where the residual $\bm{R}_0^i$ is defined as:
\begin{equation}
\bm{R}_0^i = \bm{V}_1^i-\bm{W}^i\bm{K}_1.
\label{R}
\end{equation}
$\bm{\Delta}_1^i \bm{P}$ is used to improve robustness, which is optimized by:
\begin{align}
\bm{\Delta}_1^i \bm{P} = \arg \min_{\hat{\bm{\Delta}}_1^i}\Big(&
\left\| (\bm{W^i} + \hat{\bm{\Delta}}_1^i\bm{P}) (\bm{K}_1+\bm{\epsilon}_i) - \bm{V}_1^i \right\|^{2} \nonumber \\ 
&+ \left\| \hat{\bm{\Delta}}_1^i \bm{P}\right\|^{2} \Big), \label{eq:delta1}
\end{align}
with the solution:
\begin{equation}
\bm{\Delta}_1^i\bm{P} = \bm{R}_1^i(\bm{K}_1+\bm{\epsilon}_i)\bm{P}\bigg((\bm{K}_1+\bm{\epsilon}_i)( \bm{K}_1^{\top}+\bm{\epsilon}_i^{\top})+\bm{I}\bigg)^{-1},
\end{equation}
where $\bm{\epsilon}_i$ is Gaussian noise and the residual $\bm{R}_1^i$ is given by:
\begin{equation}
\bm{R}_1^i = \bm{V}_1^i-\bm{W}^i(\bm{K}_1+\bm{\epsilon}_i).
\label{R2}
\end{equation}
In \Eref{optimize} and Eq.~\eqref{eq:delta1}, $\bm{V}_1^i = [\bm{v}_1^i \mid \bm{v}_2^i \mid \dots \mid \bm{v}_u^i]$, and $\bm{v}_j^i$ is defined as follows:
\begin{equation} 
\bm{v}_j^i = \bm{h}_j^{i} + \arg\min_{\bm{\delta}_j} (-\log \mathbb{P}_{\mathcal{F}_{\bm{W}^{i-1}}(\bm{h}_j^{i}+=\bm{\delta}_j)}\left[\bm{y}_j \mid q_j\right]), 
\label{delta}
\end{equation}
where $\bm{h}_j^i$ denotes the original hidden state of $q_j$ in layer $l$ and $\mathcal{F}_{\bm{W}^{i-1}}(\bm{h}_j^{i}+=\bm{\delta}_j)$ represents the model $\mathcal{F}_{\bm{W^{i-1}}}$ with $\bm{h}_j^{i}$ updated to 
$\bm{h}_j^{i}+\bm{\delta}_j$.

However, multi-round editing may cause the perturbation added to $\bm{W}$ to accumulate progressively, which can lead to gradient explosion and consequently degrade the effectiveness of model editing. To address this issue, we propose a stable editing strategy. Specifically, if not all QA pairs are successfully edited, $\bm{\delta}_j$ in \Eref{delta} is regularized as:
\begin{equation} 
\bm{\delta}_j \leftarrow\left\{\begin{array}{ll}
\bm{\delta}_j, & \text{}  \|\bm{\delta}_j\|_{2} \leq \varepsilon  \left\| \bm{h}_j^{i} \right\|_2 \\
\frac{\varepsilon }{\|\bm{\delta}_j\|_{2}} \bm{\delta}_j, & \|\bm{\delta}_j\|_{2}>\varepsilon  \left\| \bm{h}_j^{i} \right\|_2
\end{array}\right. ,
\end{equation}
where $\varepsilon $ is a scaling factor used to rescale $\bm{\delta}_j$ when its magnitude exceeds a reasonable range.

Conversely, if all QA pairs are successfully modified before reaching the maximum of $t$ rounds, $\bm{\delta}_j$ in \Eref{delta} is further regularized as
\begin{equation} 
\bm{\delta}_j \leftarrow\left\{\begin{array}{ll}
\bm{\delta}_j, & \text{}  \|\bm{\delta}_j\|_{2} \leq \frac{1}{t-1} \left\| \bm{h}_j^{i} \right\|_2 \\
\frac{\bm{\delta}_j}{(t-1)\|\bm{\delta}_j\|_{2}}, & \|\bm{\delta}_j\|_{2}>\frac{1}{t-1} \left\| \bm{h}_j^{i} \right\|_2
\end{array}\right. .
\end{equation}
This constraint ensures the stability of the multi-round optimization and prevents updates from exploding. In general, $\varepsilon $ is much smaller than $t$, indicating that once all QA pairs are successfully edited, the constraint on the weight updates $\bm{W}$ will be substantially tightened to ensure the stability of the editing process.
Additionally, we introduce an editing score to determine whether further updates are necessary. The score is defined as follows:
\begin{equation}
    S_i = \left\| \bm{W}_i\bm{K}_1-\bm{V}_1^i \right\|_2
\end{equation}
Given a threshold $\tau$, if $S_i<\tau$, the editing procedure is terminated early; otherwise, the optimization continues.

\subsection{Decoder}\label{decoder}
Similar to embedding watermarks, we first regenerate the sequence $\mathcal{A}=(a_1,a_2,\dots,a_u)$ using the seed and PRNG and construct MA questions $\mathcal{Q} = (q_1, q_2, \dots, q_u)$. Then, we query the suspicious LLM with the MA questions to obtain their answers $\bm{y} = (\bm{y}_1,\bm{y}_2,\dots,\bm{y}_u)$.
Let $\bm{y}_i = (\bm{y_i}^1, \bm{y_i}^2, \dots, \bm{y_i}^{m})$ denote the corresponding output for $q_i$, we design a decoding function based on lexicographical permutation theory~\cite{donald1999art} to extract the watermark $w_i$. 

Given an integer vector $\bm{b}=(a+1,a+2,\dots,a+n)$ and a permutation $\bm{\alpha} = (\alpha_1,\alpha_2,...,\alpha_m)$, the decoding function $g$ is defined as $g:\mathbb{Z}^{m} \times \mathbb{Z}^{n} \rightarrow \mathbb{N}$.
This decoding function receives two vectors $\bm{\alpha}$ and $\bm{b}$ and returns an integer. Assuming the integer output is $I$, the decoding function is defined as:
\begin{equation}
    I = g(\bm{\alpha},\bm{b})=\sum_{i=1}^{m}\left (\text{pos}(\alpha_i,\bm{b}_i) \times \frac{(n-i)!}{(n-m)!}  \right),
    \label{eq13}
\end{equation}
where $\text{pos}(\alpha_i,\bm{b}_i)$ denotes the position of $\alpha_i$ in $\bm{b}_i$, and the definition of $\bm{b}_i$ is detailed in \Eref{eq7}.

With the decoding function defined, we now describe how it is applied to extract the embedded watermark. For the MA question $q_i$ and target answer $\bm{y}_i$, the integer vector is $\bm{b}=(a_i+1,a_i+2\dots,a_i+n)$. We extract the watermark $w_i$ based on the decode function:
\begin{equation}
    w_i = \text{Binary}\big(g(\bm{y}_i,\bm{b}) \big),
    \label{eq15}
\end{equation}
where $\text{Binary}(\cdot)$ is a function that converts a decimal number to binary.

The correctness of the encoding and decoding functions is formally guaranteed by \textit{Theorem 1}.
In addition, we present the detailed algorithm and a corresponding example of the decoding function in Appendix D.

\noindent\textit{\textbf{Theorem 1.}} \textit{Given an integer vector $\bm{b}=(a,a+1,...,a+n)$, randomly select $m$ elements to form a permutation $(\alpha_1,\alpha_2,\dots,\alpha_{m)})$ and $m \le n$. The encoding function $f$ in \Eref{eq6} and the decoding function $g$ in \Eref{eq13} satisfy $g(f(I,\bm{b}),\bm{b})=I$.} 

\proof
The domain of the encoding function $f(I, \bm{b})$ is $I \in \{0, 1, \dots, \frac{n!}{(n-m)!} - 1\}$, and the range is the set of $m$ permutations $\pi_m(\bm{b})$.  
Conversely, the decoding function $g(\bm{\alpha}, \bm{b})$ takes $\bm{\alpha} \in \pi_m(\bm{b})$ and returns an integer in $\{0, 1, \dots, \frac{n!}{(n-m)!} - 1\}$.   Therefore, we can select $f(I,\bm{b})$ as input to verify whether $g(f(I,\bm{b}),\bm{b})$ is equal to $I$.

Let $f(I, \bm{b}) = \bm{\alpha} = (\alpha_1, \alpha_2, \dots, \alpha_{m})$, we have:
\begin{align}
g(f(I,\bm{b}),\bm{b}) 
    =\sum_{i=1}^{m}\left ( \text{pos}(\alpha_i,\bm{b}_i) \times \frac{(n-i)!}{(n-m)!} \right ).
\end{align}
Since $\text{pos}(\alpha_i,\bm{b}_i)=\left \lfloor \frac{I_i}{\frac{(n-i)!}{(n-m)!}} \right \rfloor$, we have:
\begin{equation}
    g(f(I,\bm{b}),\bm{b})= \sum_{i=1}^{m} \left ( \left \lfloor \frac{I_i}{\frac{(n-i)!}{(n-m)!}} \right \rfloor \times \frac{(n-i)!}{(n-m)!} \right ).
    \label{eq25}
\end{equation}
Using $\left\lfloor \frac{d_1}{d_2} \right\rfloor = \frac{d_1 - (d_1 \bmod d_2)}{d_2}$, we simplify~\Eref{eq25}:
\begin{equation}
    g(f(I, \bm{b}), \bm{b}) = \sum_{i=1}^{m} \left( I_i - I_i \bmod \frac{(n-i)!}{(n-m)!} \right).
    \label{eq26}
\end{equation}
Since $I_{i+1} = I_i \bmod \frac{(n-i)!}{(n-m)!}$, we simplify~\Eref{eq26}:
\begin{align}
    g(f(I,\bm{b}),\bm{b})& = \sum_{i=1}^{m}(I_i-I_{i+1}) = I_1-I_{m+1}.
\end{align}
As $I_{m+1}=I_{m} \bmod \frac{(n-m)!}{(n-m)!}=0$, we have:
\begin{align}
    g(f(I,\bm{b}),\bm{b})&= I_1=I.
\end{align}
Therefore, we can prove that $g(f(I,\bm{b}),\bm{b})=I$. Hence, the decoding function precisely inverts the encoding function, completing the proof.
\qedhere

\begin{table*}[t]
\centering
\caption{The performance of EditMark and baselines on watermark extraction success rate and embedding time.}
\label{main}
\scalebox{0.95}{
\begin{tblr}{
  cells = {c},
  cell{1}{1} = {r=2}{},
  cell{1}{2} = {r=2}{},
  cell{1}{3} = {c=2}{},
  cell{1}{6} = {c=2}{},
  cell{1}{9} = {c=2}{},
  cell{1}{12} = {c=2}{},
  cell{1}{15} = {c=2}{},
  hline{1,7} = {-}{0.08em},
  hline{2} = {3-4,6-7,9-10,12-13,15-16}{},
  hline{3} = {-}{},
}
Method     & {Watermark\\Capacity}        & GPT2-XL        &                &  & GPT-J-6B       &                &  & LLaMA-3-8B     &                &  & Baichuan-7B    &                &  & Qwen-7B        &               \\
           &                 & ESR            & ET             &  & ESR            & ET             &  & ESR            & ET             &  & ESR            & ET             &  & ESR            & ET            \\
Backdoor~\cite{li-etal-2023-watermarking,xu2024instructional} & 1-bit           & 57.2\%         & 3399.8s        &  & 83.5\%         & 3386.3s        &  & 2.2\%          & 5198.2s        &  & 32.2\%         & 3924.4s        &  & 7.2\%          & 4367.8s       \\
KIMark~\cite{li2024}    & \textbf{32-bit} & \textbf{100\%} & 5563.8s        &  & \textbf{100\%} & 5803.4s        &  & \textbf{100\%} & 9002.7s        &  & \textbf{100\%} & 6560.8s        &  & 99.1\%         & 7450.2s       \\
BadEdit~\cite{libadedit}   & 1-bit           & 88.5\%         & 110.8s         &  & 99.7\%         & 68.2s          &  & 93.0\%         & 149.5s         &  & 92.7\%         & 114.5s         &  & 98.4\%         & 37.3s         \\
EditMark   & \textbf{32-bit} & \textbf{100\%} & \textbf{12.2s} &  & \textbf{100\%} & \textbf{12.9s} &  & \textbf{100\%} & \textbf{12.0s} &  & \textbf{100\%} & \textbf{9.47s} &  & \textbf{100\%} & \textbf{9.9s} 
\end{tblr}}
\end{table*}

\begin{table*}[t]
\centering
\caption{The performance of EditMark on different watermark capacities.}
\label{main2}
\scalebox{0.9}{
\begin{tblr}{
  cells = {c},
  cell{1}{1} = {r=2}{},
  cell{1}{2} = {c=3}{},
  cell{1}{6} = {c=3}{},
  cell{1}{10} = {c=3}{},
  cell{1}{14} = {c=3}{},
  cell{1}{18} = {c=3}{},
  hline{1,5} = {-}{0.08em},
  hline{2} = {2-4,6-8,10-12,14-16,18-20}{},
  hline{3} = {-}{},
}
Metric & GPT2-XL &        &         &  & GPT-J-6B &        &         &  & LLaMA-3-8B &        &         &  & Baichuan-7B &        &         &  & Qwen-7B &        &         \\
       & 32-bit  & 64-bit & 128-bit &  & 32-bit   & 64-bit & 128-bit &  & 32-bit     & 64-bit & 128-bit &  & 32-bit      & 64-bit & 128-bit &  & 32-bit  & 64-bit & 128-bit \\
ESR    & 100\%   & 100\%  & 100\%   &  & 100\%    & 100\%  & 100\%   &  & 100\%      & 100\%  & 100\%   &  & 100\%       & 100\%  & 100\%   &  & 100\%   & 100\%  & 100.0\%  \\
ET     & 12.0s   & 30.6s  & 52.7s   &  & 14.1s    & 20.2s  & 36.5s   &  & 12.0s      & 22.5s  & 61.5s   &  & 9.4s        & 16.8s  & 32.8s   &  & 9.9s    & 14.9s  & 27.1s   
\end{tblr}}
\end{table*}

\section{Experiments}

\subsection{Experimental Settings}
\noindent\textbf{Model:} To evaluate the effectiveness of the EditMark framework, we evaluate our method on five common large language models: GPT2-XL~\cite{radford2019language}, GPT-J-6B~\cite{mesh-transformer-jax}, LLaMA-3-8B~\cite{llama3modelcard}, Baichuan-7B~\cite{baichuan2023baichuan2}, and Qwen-7B~\cite{bai2023qwen}.

\noindent\textbf{Model Editing Technique:} We select \textit{AlphaEdit}~\cite{fang2024alphaedit} as the model editing technique to embed a watermark. 

\noindent\textbf{Metrics:} We employ two principal metrics to evaluate EditMark: extraction success rate (ESR) and embedding time (ET). ESR is the ratio of MA questions whose embedded watermark is correctly extracted to the total number of questions. ET is measured as the average editing time per watermark embedding process, excluding model loading and inference time.

\noindent\textbf{Hyperparameters:} Unless otherwise stated, we set $\beta=32$, $m=5$ and $n=89$. For each LLM, the maximum number of edit epochs is $t=3$, $\lambda=0.3$, $\tau=0.5$. 

\noindent\textbf{Baseline:} We consider three well-established watermarking techniques: the watermarking method based on backdoor~\cite{xu2024instructional,li-etal-2023-watermarking}, knowledge injection (KIMark)~\cite{li2024}, and BadEdit~\cite{libadedit}. While BadEdit was originally proposed for injecting backdoors in LLMs,  we adapt it to the watermarking scenario. The details of the baselines are provided in Appendix A.

\noindent\textbf{Implementation Details:} The MLP layer we edit is ``15'' for LLaMA-3-8B, ``17'' for GPT2-XL and GPT-J-6B, and ``14'' for Qwen-7B and ``Baichuan-7B''. All experiments were conducted on four NVIDIA RTX 4090 GPUs (24GB each), and the models were loaded in float16 precision. We try to embed 32-bit, 64-bit, and 128-bit watermarks into LLM. The MA question templates we used are detailed in Table~\ref{template}. Under different watermark capacities, we set the seed 1-20 to conduct experiments to eliminate the impact of randomness.

\subsection{Effectiveness}
An effective watermarking method should achieve a high watermark extraction success rate while minimizing the watermark embedding time. To evaluate the effectiveness of EditMark, we calculate both the ESR and the time cost for embedding the watermark across different LLMs.

As shown in Table~\ref{main}, EditMark achieves a 100\% ESR and requires less than 20 seconds to embed a 32-bit watermark for all LLMs evaluated. In particular, the average embedding time for Baichuan-7B and Qwen-7B is under 10 seconds, which demonstrates the high efficiency of EditMark. 
In addition, we compare EditMark against several baseline methods, including Backdoor, KIMark, and BadEdit. The results show that EditMark outperforms the baselines in terms of ESR and watermark embedding time, highlighting the effectiveness and
efficiency of our watermarking scheme. 
Furthermore, we assess the ESR and time cost of EditMark when embedding watermarks of larger capacities. As presented in Table~\ref{main2},
the embedding time remains under 2 minutes with a 100\% ESR for all evaluated LLMs when embedding a 128-bit watermark, validating that EditMark remains effective and efficient for larger watermark capacities.



\subsection{Fidelity}
The fidelity of a watermarking method refers to its ability to embed a watermark with minimal impact on the performance of the original LLM. In other words, the watermarked LLM should perform similarly to the original model.
To evaluate the performance of watermarked LLMs, we assess their performance on four benchmark tasks: MMLU~\cite{MMLU}, BLIMP~\cite{blimp}, TruthfulQA~\cite{truth}, and GLUE~\cite{glue}. Among these, TruthfulQA, MMLU, and GLUE evaluate the model’s general knowledge, while BLIMP focuses on the model's language-understanding capabilities. For each LLM, we embed a 32-bit and 128-bit watermark using seeds 1-5, respectively, to obtain ten watermarked LLMs, and then calculate their average performance on these benchmarks. 

As shown in Table~\ref{table3}, when embedding a 32-bit or 128-bit watermark, the average performance of the watermarked LLMs remains close to that of the original LLMs across multiple benchmarks. Moreover, even with an increased watermark capacity, the performance of the watermarked model does not degrade significantly. These results indicate that EditMark is capable of embedding watermarks while preserving the performance, thus demonstrating its fidelity.

\begin{table*}[t]
\centering
\caption{Evaluation of watermark fidelity across multiple benchmarks. ``Original'' denotes the unmodified model, while ``32-bit'' and ``128-bit'' indicate watermarked models with different capacities.}
\label{table3}
\scalebox{0.95}{
\begin{tblr}{
  row{even} = {c},
  row{3} = {c},
  row{5} = {c},
  row{7} = {c},
  cell{1}{1} = {r=2}{c},
  cell{1}{2} = {c=3}{c},
  cell{1}{5} = {c},
  cell{1}{6} = {c=3}{c},
  cell{1}{9} = {c},
  cell{1}{10} = {c=3}{c},
  cell{1}{14} = {c=3}{c},
  hline{1,9} = {-}{0.08em},
  hline{2} = {2-4,6-8,10-12,14-16}{},
  hline{3} = {1-12,14-16}{},
}
Models      & BLIMP~ &                 &                 &  & MMLU~  &                 &                 &  & TruthfulQA~ &                 &         &  & GLUE~  &        &                 \\
            & Original   & 32-bit          & 128-bit         &  & Original   & 32-bit          & 128-bit         &  & Original        & 32-bit          & 128-bit &  & Original   & 32-bit & 128-bit         \\
GPT2-XL     & 81.3\% & 81.2\%          & 81.2\%          &  & 27.1\% & 27.6\%          & 27.6\%          &  & 37.9\%      & 38.0\%          & 38.0\%  &  & 48.9\% & 49.4\% & 49.3\%          \\
GPT-J-6B    & 81.1\% & 81.3\%          & 81.3\%          &  & 27.5\% & 28.0\%          & 27.6\%          &  & 36.4\%      & 37.2\%          & 37.2\%  &  & 48.5\% & 49.1\% & 50.0\%          \\
LLaMA-3-8B  & 74.3\% & 74.3\%          & 74.3\%          &  & 67.3\% & 67.3\%          & 67.0\%          &  & 42.3\%      & 42.3\%          & 42.3\%  &  & 57.5\% & 57.4\% & 57.6\%          \\
Baichuan-7B & 81.3\% & 81.3\%          & 81.4\%          &  & 40.7\% & 41.0\%          & 41.1\%          &  & 38.0\%      & 38.1\%          & 38.0\%  &  & 47.9\% & 48.6\% & 48.6\%          \\
Qwen-7B     & 81.2\% & 81.6\%          & 81.3\%          &  & 58.0\% & 58.3\%          & 57.7\%          &  & 48.7\%      & 48.9\%          & 48.9\%  &  & 70.0\% & 70.1\% & 69.9\%          \\
Average     & 79.8\% & \textbf{79.9\%} & \textbf{79.9\%} &  & 44.1\% & \textbf{44.4\%} & \textbf{44.4\%} &  & 40.7\%      & \textbf{41.0\%} & 40.9\%  &  & 54.6\% & 54.9\% & \textbf{55.1\%} 
\end{tblr}}
\end{table*}

\subsection{Robustness}
In the open-source scenario of watermarking LLMs, robustness is crucial for watermarking methods. This is because the attacker may attempt to use various attack strategies to remove the embedded watermark. An excellent watermarking method should be robust against classic attack methods, e.g., fine-tuning attacks, quantization attacks, random noise attacks, and pruning attacks. 

To evaluate the robustness of EditMark, we select six classic attack methods to test the robustness of the watermarked LLMs. These attacks include fine-tuning attacks, quantization attacks, random noise attacks, pruning attacks, model editing attacks, and adaptive attacks. In the following sections, we will describe these six attack scenarios in detail and evaluate how EditMark performs under each of these attacks.
In our evaluation, for each LLM, we embed a 128-bit watermark using seeds 1-5 to obtain five corresponding watermarked LLMs.

\subsubsection{Fine-tuning Attacks} 
\begin{figure}[h]
\centering
    \centering
    \includegraphics[width=0.48\textwidth]{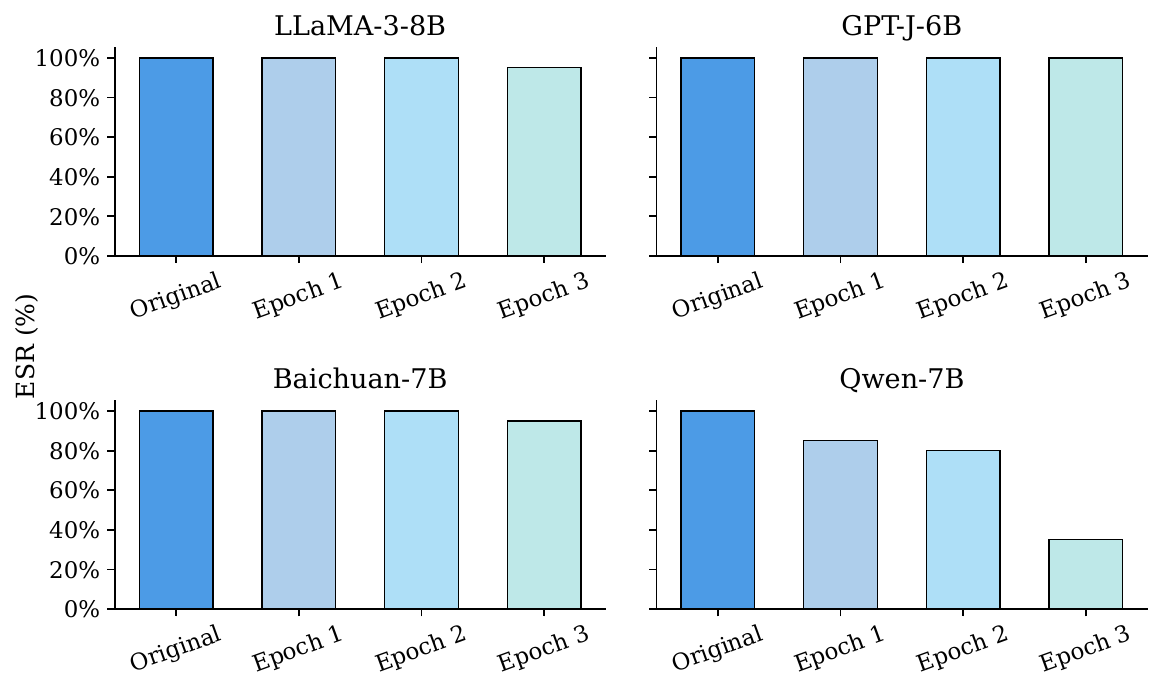}
    \caption{ESR of original watermarked LLMs and the attacked LLMs under model fine-tuning attacks.}
    \label{fig:fig4}
\end{figure}

Recently, LoRA fine-tuning~\cite{lora} has gained widespread adoption in various tasks due to its low cost and minimal impact on the fine-tuned LLM. However, LoRA fine-tuning also presents an effective method for watermark removal. Attackers can modify the parameters of a watermarked LLM through LoRA fine-tuning, potentially removing the watermark while preserving the model's overall performance.
To evaluate the robustness of EditMark against the model fine-tuning attack, we fine-tune the LLMs on a subset of the C4\footnote{\url{https://huggingface.co/datasets/allenai/c4}} dataset, which includes 2000 texts. 
As shown in Figure \ref{fig:fig4}, we anticipate that the ESR of most LLMs slightly decreases as the number of fine-tuning epochs increases. After just one epoch of fine-tuning, the watermark remains largely intact. Although Qwen-7B exhibits a noticeable decline in ESR after three epochs, the other models maintain an extraction success rate exceeding 90\%. We attribute this variation in robustness across models to differences in perturbation strength introduced during editing.
Overall, these results indicate that our watermarking method exhibits a considerable resilience against fine-tuning attacks.

\subsubsection{Quantization Attacks}
\begin{figure}[h]
\centering
    \centering
    \includegraphics[width=0.48\textwidth]{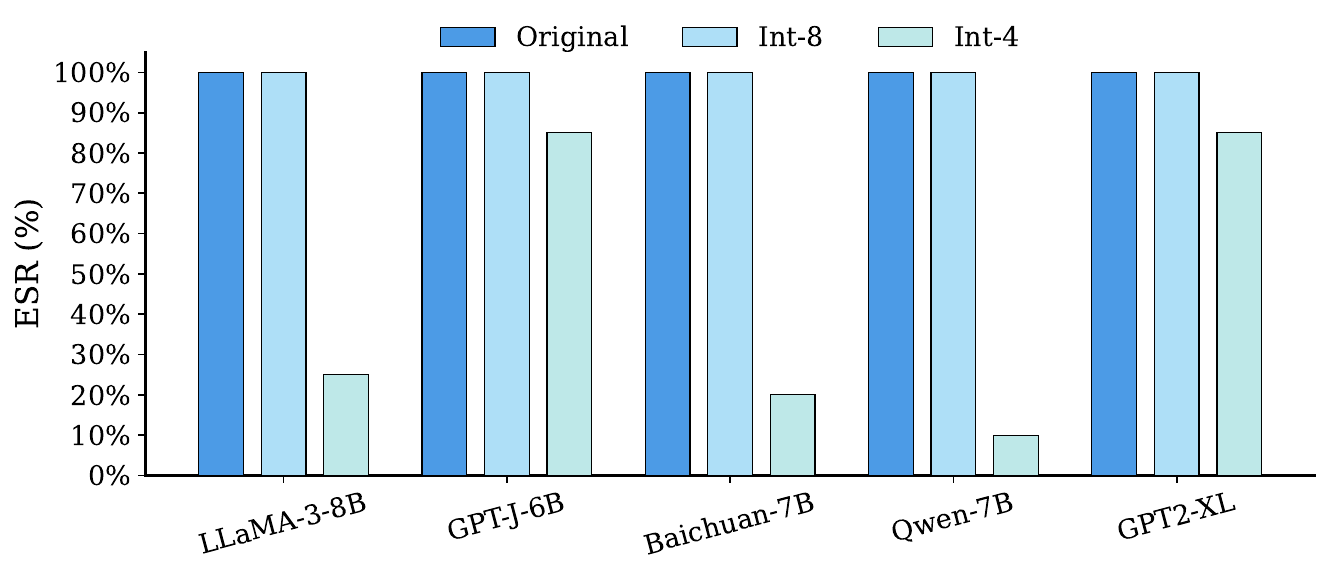}
    \caption{The ESR of original watermarked LLMs and the attacked LLM by quantification. }
    \label{fig:fig5}
\end{figure}

Quantization is a common technique used to reduce the size and computational cost of LLMs by lowering the numerical precision of model parameters, typically without significantly affecting model performance. However, this process may also remove or degrade the embedded watermark, a vulnerability known as a quantization attack.
In this study, we evaluate the robustness of EditMark against quantization attacks using two widely adopted quantization schemes: Int-4 and Int-8. We apply both methods to quantize the watermarked LLMs and assess their impact on ESR.
As shown in Figure \ref{fig:fig5}, when the watermarked LLMs are quantized using Int-8, the ESR remains 100\% across all evaluated models, indicating that the watermark cannot be fully removed by Int-8 quantization. Although Int-4 quantization can partially reduce the ESR, it will also significantly degrade the overall performance of the affected LLMs. Therefore, these results indicate that EditMark possesses a certain degree of robustness against model quantization attacks.

\begin{figure*}[t]
\centering
    \centering
    \includegraphics[width=0.95\textwidth]{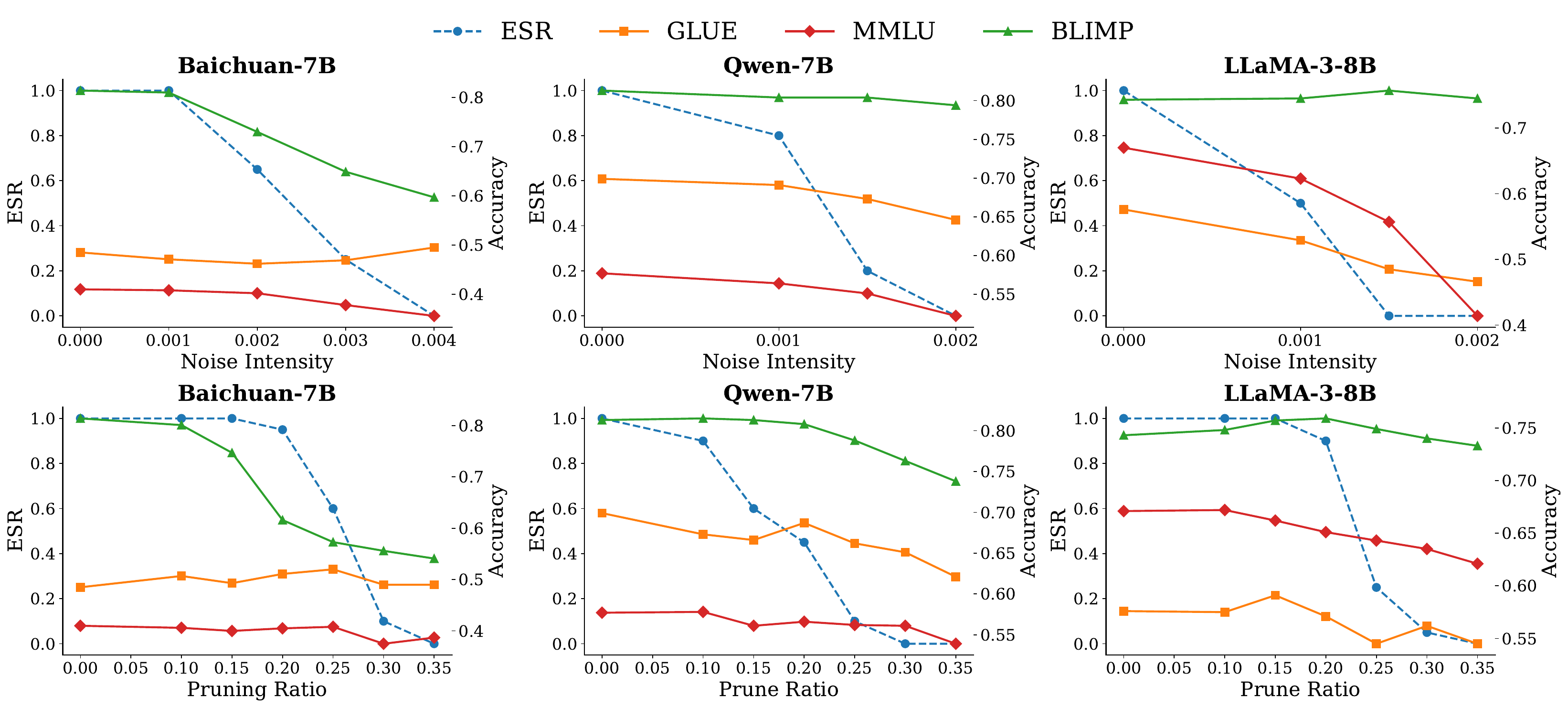}
    \caption{The ESR and performance of the attacked LLMs with noise attacks and pruning attacks under different noise intensities and pruning ratios.}
    \label{fig:fig6}
\end{figure*}

\subsubsection{Noise Attacks} Noise attacks aim to destroy embedded watermarks by injecting random perturbations into the weights of the watermarked LLM. In this paper, we simulate such attacks by adding Gaussian noise of varying intensity to all attention and MLP layers of the watermarked models. The noise intensity is defined as the standard deviation of the Gaussian distribution.
As shown in Figure~\ref{fig:fig6}, as the noise intensity increases, the ESR gradually decreases. However, this degradation in ESR is accompanied by a noticeable drop in the model's performance. For instance, Baichuan-7B shows a 31.3\% performance drop on the BLIMP dataset, Qwen-7B experiences a 5.3\% decline on GLUE, and LLaMA-3-8B suffers a 11.2\% decrease on MMLU, compared with the original models.
These results indicate that removing watermarks in watermarked LLMs with random noise requires sacrificing the performance of the watermarked LLMs, which is detrimental to the attacker. In summary, it is challenging for an attacker to completely remove the watermark while maintaining the performance of the targeted LLM, suggesting that EditMark exhibits a certain degree of robustness against random noise attacks.

\subsubsection{Pruning Attacks}
Model pruning aims to reduce the computational and storage overhead of an LLM while maintaining its performance by removing redundant or less contributing components (such as weights and neurons) in a neural network. Unfortunately, attackers can also utilize the model pruning technique to remove the watermark embedded in watermarked LLMs, a technique known as a pruning attack. To evaluate the robustness of EditMark against pruning attacks, we adopt unstructured pruning, where a specified proportion of the weights with the smallest magnitudes across the MLP layers are set to zero.
As shown in Figure \ref{fig:fig6}, although the ESR gradually decreases as the pruning ratio increases, the performance of the pruned model will also decrease. For example, Baichuan-7B shows a 27.3\% performance drop on the BLIMP dataset, Qwen-7B experiences a 4.8\% decline on GLUE, and LLaMA-3-8B suffers a 5.0\% decrease on MMLU, compared to the original models. Similar to random noise attacks, pruning can remove the watermark, but it also results in a marked degradation of model performance on certain benchmarks. This demonstrates that our watermarking method maintains a degree of robustness against pruning attacks.

\subsubsection{Model Editing Attacks}
Model editing attacks remove the watermark by editing the parameters of the watermarked LLM. To evaluate the robustness of EditMark against model editing attacks, we edit the watermarked LLM on the \textit{Counterfact} dataset~\cite{rome} and calculate the ESR after the attack. Specifically, we select 10-50 samples from the dataset and apply edits to the watermarked models. In this evaluation, we assume a strong adversary who knows the exact MLP layers where the watermark was embedded and directly edits those same layers to attempt watermark removal.
As shown in Table~\ref{table:edit-attack}, the ESR remains above 95\% across all attacked models under this attack. This demonstrates that EditMark maintains a high ESR even under direct model editing attacks, highlighting its robustness. Furthermore, in realistic scenarios, an attacker may not know which specific layer contains the watermark, significantly increasing the difficulty and cost of a successful removal attack.

\begin{table}[h]
\centering
\caption{ESR of watermarked LLMs under model editing attacks.}
\label{table:edit-attack}
\begin{tblr}{
  cells = {c},
  cell{1}{1} = {r=2}{},
  cell{1}{2} = {c=6}{},
  hline{1,8} = {-}{0.08em},
  hline{2} = {2-7}{},
  hline{3} = {-}{},
}
Model       & Number of Edit Case &       &       &        &        &        \\
            & Original            & 10    & 20    & 30     & 40     & 50     \\
GPT2-XL     & 100\%               & 100\% & 100\% & 100\%  & 100\%  & 100\%  \\
GPT-J-6B    & 100\%               & 100\% & 100\% & 95.0\% & 100\%  & 95.0\% \\
LLaMA-3-8B  & 100\%               & 100\% & 100\% & 100\%  & 95.0\% & 95.0\% \\
Baichuan-7B & 100\%               & 100\% & 100\% & 100\%  & 100\%  & 100\%  \\
Qwen-7B     & 100\%               & 100\% & 100\% & 100\%  & 100\%  & 100\%  
\end{tblr}
\end{table}

\subsubsection{Adaptive Attacks}\label{ada}
An adaptive attacker possesses full or partial knowledge of the watermarking algorithm. In such cases, the attacker may embed a new watermark into an already watermarked LLM, thereby overwriting the original one. This type of attack, often referred to as an overwrite or forgery attack, poses a serious threat to the integrity of the embedded watermark.  
To evaluate the robustness of EditMark under such conditions, we measure the ESR of watermarked LLMs against adaptive attacks. Specifically, we assume that the attacker is aware of the model editing technique and the target MLP layer used for watermark embedding, but does not know the subject $s$ in our MA question template. Depending on the attacker's knowledge of the MA question template, we consider two scenarios:  
\begin{itemize}
    \item  Scenario A: The attacker knows the MA question template except for the subject.  
    \item  Scenario B: The attacker has no knowledge of the MA question template.  
\end{itemize}
The examples of MA question templates used in both scenarios are provided in Table~\ref{table:adaptive-attack}.  
As shown in Table~\ref{adaptive},  the ESR drops significantly in scenario A, indicating that when attackers have precise knowledge of the MA question template, adaptive attacks can effectively erase the watermark. However, in the more practical scenario B, most models still retain an ESR above 30\%, demonstrating that EditMark maintains a certain degree of robustness against adaptive attacks.

\begin{table}[h]
\centering
\caption{The example of MA question templates for adaptive attack.}
\label{table:adaptive-attack}
\scalebox{0.78}{
\begin{tblr}{
  cells = {c},
  cell{2}{1} = {r=2}{},
  cell{4}{1} = {r=2}{},
  hline{1,6} = {-}{0.08em},
  hline{2} = {-}{},
}
Scenario & Template                                               \\
A        & For the inequality $\frac1{a'}<\frac1{x}<\frac1{a}$, 5 random integer solutions are $x=$ \\
         & For the inequality $a+1<y+1<a'+1$, 5 random integer solutions are $y=$ \\
B        & 5 random integer solutions for the inequality $\frac1{a'}<\frac1{x}<\frac1{a}$ are $x =$  \\
         & 5 random integer solutions for the inequality $a+1<y+1<a'+1$ are $y =$  
\end{tblr}}
\end{table}

\begin{table}[h]
\centering
\caption{The ESR of watermarked LLMs under adaptive attack.}
\label{adaptive}
\begin{tblr}{
  cells = {c},
  hline{1,7} = {-}{0.08em},
  hline{2} = {-}{},
}
Model       & Original & Scenario A & Scenario B \\
GPT2-XL     & 100\%    & 40\%      & 52.0\%      \\
GPT-J-6B    & 100\%    & 26.0\%    & 44.0\%     \\
LLaMA-3-8B  & 100\%    & 10.0\%    & 33.0\%     \\
Baichuan-7B & 100\%    & 3.0\%     & 4.0\%     \\
Qwen-7B     & 100\%    & 29.0\%    & 74.0\%      
\end{tblr}
\end{table}

\subsection{Stealthiness}
Stealthiness is a crucial property for watermarking methods during watermark extraction. If the watermarked text appears unnatural or suspicious, an attacker may identify and remove the watermark through adaptive attacks. Therefore, the watermarked outputs should be both logical and indistinguishable from natural text.
To assess the stealthiness of EditMark, we compute the perplexity (\textit{PPL}) of watermarked texts generated by EditMark and compare it to that of natural texts sampled from the C4 and AGNews~\footnote{\url{https://huggingface.co/datasets/fancyzhx/ag_news}} datasets. Specifically, we randomly generate 1,000 watermarked texts and sample 2,000 natural texts from each dataset. Since the watermarked texts are relatively short (average token length: 55), we truncate the natural texts to maximum token lengths of 64 and 128 to ensure a fair comparison.

As shown in Table~\ref{ppl}, the average \textit{PPL} of the watermarked texts produced by EditMark is close to that of natural texts, indicating that they are linguistically natural and difficult to distinguish using perplexity-based methods. For comparison, we also report the \textit{PPL} of watermarked texts generated by BadEdit, which is significantly higher than that of natural texts. This suggests that BadEdit produces text that is noticeably less fluent, reducing its stealthiness during inference.
In summary, the results demonstrate that EditMark maintains high stealthiness, as the generated watermarked texts remain natural and difficult to detect through standard language modeling metrics such as \textit{PPL}.

\begin{figure*}[t]
\centering
    \centering
    \includegraphics[width=\textwidth]{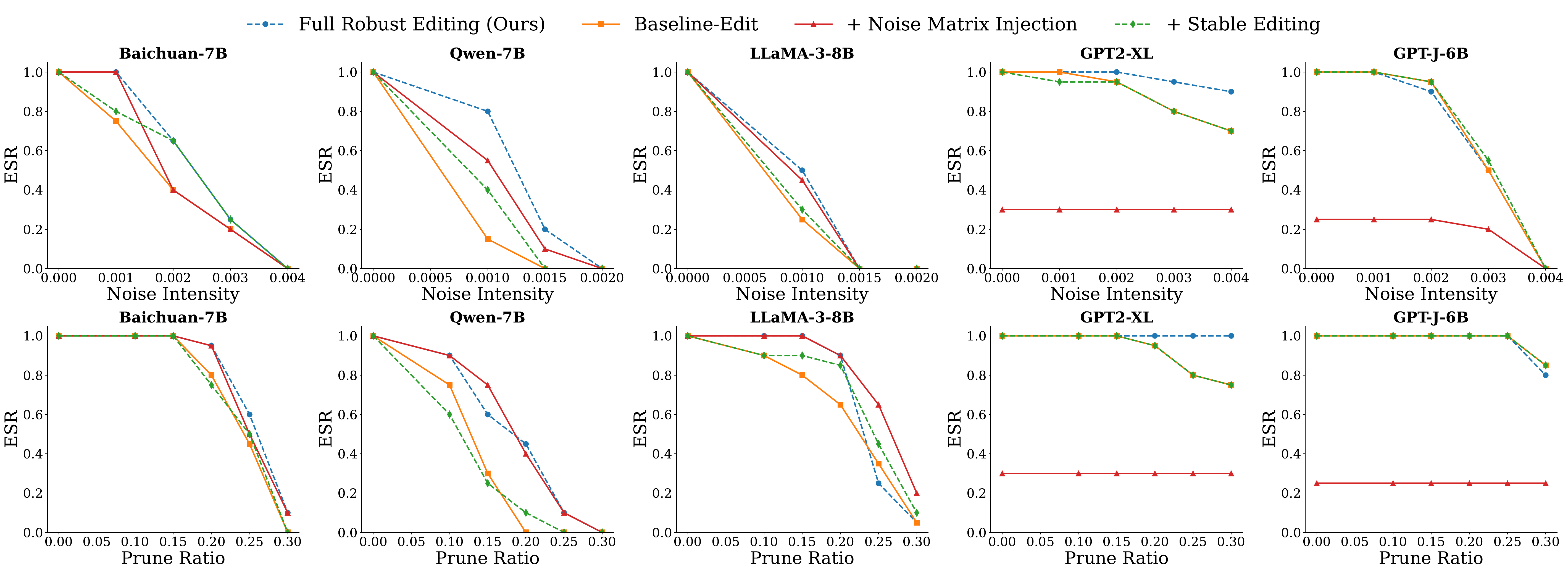}
    \caption{Comparison of our robust editing method and the baseline editing method in terms of robustness against random noise and model pruning attacks.}
    \label{fig:ablation_robust}
\end{figure*}

\begin{table}[h]
\centering
\caption{The average \textit{PPL} of watermarked texts and natural texts. 64 and 128 represent the maximum token length of natural texts to cut off.}
\label{ppl}
\scalebox{0.9}{
\begin{tblr}{
  cells = {c},
  cell{1}{1} = {r=2}{},
  cell{1}{2} = {c=2}{},
  cell{1}{5} = {c=2}{},
  cell{1}{7} = {r=2}{},
  cell{1}{8} = {r=2}{},
  hline{1,7} = {-}{0.08em},
  hline{2} = {2-3,5-6}{},
  hline{3} = {-}{},
}
Model       & C4    &       &  & AGNews &       & BadEdit & EditMark \\
            & 64    & 128   &  & 64     & 128   &         &          \\
LLaMA-3-8B  & 30.44 & 23.6  &  & 34.09  & 34.09 & 137.05   & 27.01    \\
Baichuan-7B & 14.84 & 12.27 &  & 17.37  & 16.89 & 72.32   & 10.20    \\
Qwen-7B     & 22.21 & 21.98 &  & 22.89  & 18.19 & 62.48   & 10.34    \\
LLaMA-13B   & 12.15 & 9.91  &  & 12.29  & 11.84 & 72.86   & 11.18    
\end{tblr}}
\end{table}

\subsection{Ablation Experiments}

\subsubsection{Multi-round Stable Editing Strategy}\label{Multi-round}
To validate the effectiveness of the proposed adaptive multi-round stable editing strategy, we conduct an ablation study by comparing it with the original model editing technique \textit{AlphaEdit}. Specifically, for the standard \textit{AlphaEdit}, the model parameters are updated only once during the optimization objective. In contrast, the adaptive multi-round stable editing strategy is performed iteratively until the residual on the target keys falls below a predefined threshold.

The results of two editing strategies are reported in Table~\ref{tab:ablation-multiround}. 
We observe that adaptive multi-round editing consistently outperforms single-shot editing across all settings, demonstrating its importance in improving the reliability of watermark embedding. Notably, the performance gain becomes more pronounced when the watermark embedding task involves a longer length of watermark message, which further highlights the necessity of iterative refinement.

\begin{table}[h]
\centering
\caption{The ESR of watermarked LLMs under different editing strategies.}
\label{tab:ablation-multiround}
\begin{tblr}{
  cells = {c},
  cell{1}{1} = {r=2}{},
  cell{1}{2} = {c=3}{},
  cell{1}{5} = {c=3}{},
  hline{1,8} = {-}{0.08em},
  hline{2} = {2-7}{},
  hline{3} = {-}{},
}
Model       & Original Editing Technique &        &            & Ours Editing Technique &        &         \\
            & 32-bit              & 64-bit & 128-bit    & 32-bit              & 64-bit & 128-bit \\
GPT2-XL     & 100\%               & 82.5\% & 90.0\%     & 100\%               & 100\%  & 100\%   \\
GPT-J-6B    & 0\%                 & 0\%    & 2.5\%     & 100\%               & 100\%  & 100\%   \\
Qwen-7B     & 100\%               & 90.0\% & 86.2\%     & 100\%               & 100\%  & 100\%   \\
Baichuan-7B & 100\%               & 95.0\% & 55.0\%     & 100\%               & 100\%  & 100\%   \\
LLaMA-3-8b  & 95.0\%              & 100\%  & 73.7\% & 100\%               & 100\%  & 100\%   
\end{tblr}
\end{table}

\subsubsection{Robust Editng Strategy}\label{robust-editing}
Beyond improving the editing success rate, another critical requirement for watermark embedding is robustness against adversarial manipulations. To this end, our framework incorporates two key techniques: 1) noise injection, where a Gaussian noise matrix is added to the key representations before computing perturbations, and 2) stable editing, which prevents excessive updates by introducing an editing score for early stopping and by regularizing the magnitude of $\bm{\delta}_j$.
To evaluate their contributions, we conduct an ablation study under four settings:
\begin{itemize}
    \item Baseline-Edit (w/o Robust Editing): Standard multi-round editing without noise injection or stable editing.
    \item + Noise Matrix Injection: Incorporating Gaussian noise into the keys while keeping the editing update unchanged.
    \item + Stable Editing: Employing the stable editing and update regularization without noise matrix injection.
    \item Full Robust Editing (ours): Combining both noise matrix injection and stable editing.
\end{itemize}

The results in Figure~\ref{fig:ablation_robust} show that noise matrix injection substantially improves the robustness of watermark embedding, but without Stable Editing, it may trigger gradient explosion and cause editing failures (e.g., on GPT-J-6B and GPT2-XL). Stable Editing, on the other hand, stabilizes the optimization by preventing excessive updates, yet its robustness improvement alone is limited. Therefore, the combination of noise matrix injection and Stable Editing is essential, as it jointly improves both robustness and the stability and reliability of the editing process.

\subsubsection{Temperature}
The \textit{temperature} hyperparameter in the inference of LLMs controls the randomness of the output. A lower \textit{temperature} (e.g., 0.2) makes the LLM more deterministic, favoring high-probability responses. In contrast, a higher \textit{temperature}  (e.g., 1.0) increases diversity and creativity during inference. To evaluate the impact of \textit{temperature} on watermark extraction, we compute the average ESR of watermarked LLMs under different settings of \textit{temperature}. For each setting, watermark extraction is performed under six independent experiments per model to reduce the effect of sampling randomness.
As shown in Table~\ref{temperature}, EditMark achieves an ESR above 100\% across all models at a \textit{temperature} of 0.5, indicating strong performance under typical inference conditions. Even at a \textit{temperature } of 1.0, EditMark also maintains a 100\% ESR for most models, demonstrating that EditMark is also feasible on a higher sampling randomness during inference.

\begin{table}[h]
\centering
\caption{The average ESR under different temperatures.}
\label{temperature}
\begin{tblr}{
  cells = {c},
  cell{1}{1} = {r=2}{},
  cell{1}{2} = {c=4}{},
  hline{1,8} = {-}{0.08em},
  hline{2} = {2-5}{},
  hline{3} = {-}{},
}
Model       & Temperature &         &         &        \\
            & 0.0         & 0.25    & 0.5     & 1.0    \\
LLaMA-3-8b  & 100.0\%     & 100.0\% & 100.0\%  & 100.0\% \\
GPT2-XL     & 100.0\%     & 100.0\% & 100.0\% & 100.0\% \\
Qwen-7B     & 100.0\%     & 100.0\% & 100.0\% & 100.0\% \\
Baichuan-7B & 100.0\%     & 100.0\% & 100.0\%  & 100.0\% \\
GPT-J-6B    & 100.0\%     & 100.0\% & 100.0\% & 65.9\% 
\end{tblr}
\end{table}

\subsubsection{Model Size}
To eliminate the impact of model size on watermarking LLM, we select the LLM with a larger size: Qwen-14B. The MLP layer for the LLM we edit is `14'. Due to the limitation of computing resources, we used an RTX A6000 to edit the LLM.
As shown in Table \ref{size}, the average ESR of larger LLMs is 100\% when embedding 32-bit, 64-bit, and 128-bit watermarks, which demonstrates that EditMark is still effective for larger LLMs. In addition, although the watermark embedding time increases with the model size, it is still significantly less than the time required for model fine-tuning, which demonstrates the efficiency of EditMark.

\begin{figure*}[t]
\centering
    \centering
    \includegraphics[width=\textwidth]{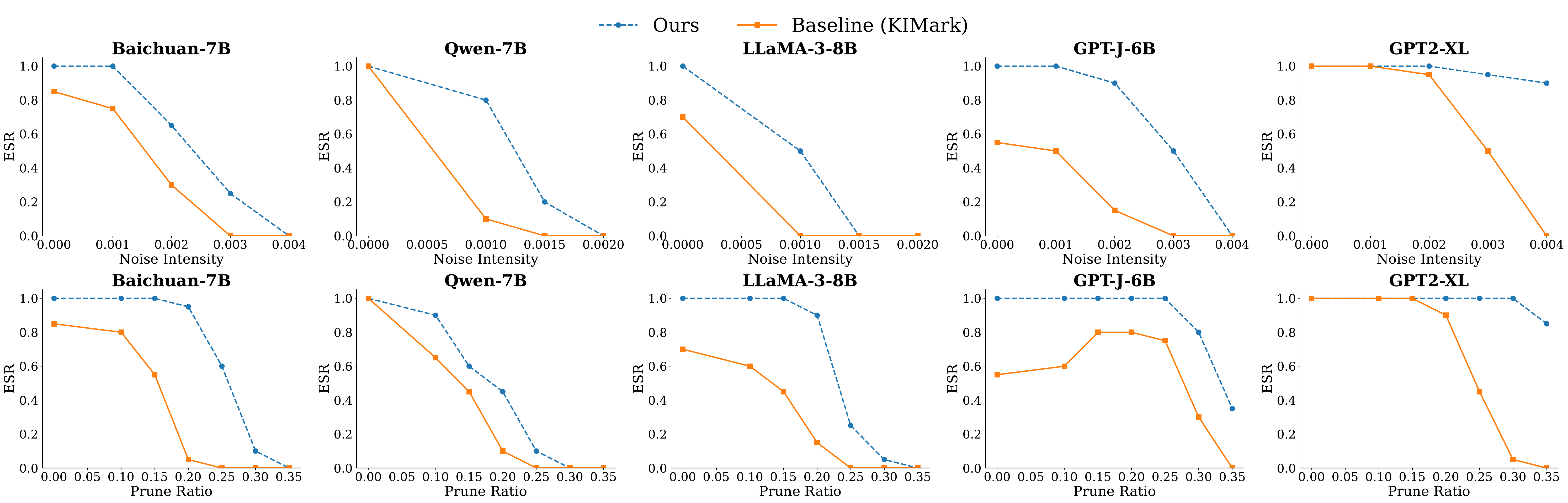}
    \caption{Comparison of our method and the baseline in terms of robustness against random noise and model pruning attacks.}
    \label{fig:noise_compare}
\end{figure*}

\begin{table}[h]
\centering
\caption{The ESR and ET of watermarked LLMs with a larger size.}
\label{size}
\begin{tblr}{
  cells = {c},
  cell{2}{1} = {r=3}{},
  hline{1,5} = {-}{0.08em},
  hline{2} = {-}{},
}
Model     & Capacity & ESR     & ET     \\
Qwen-14B  & 32-bit   & 100.0\% & 18.2s  \\
          & 64-bit   & 100.0\% & 30.3s  \\
          & 128-bit  & 100.0\% & 63.6s 
\end{tblr}
\end{table}

\subsubsection{Generation Strategy of MA Question}\label{Generation Strategy}
To mitigate the entanglement issue in model editing, we adopt a diverse MA question strategy to generate multiple-answer questions. To validate the effectiveness of this strategy, we compare it against a baseline where only a single template is used to generate MA questions, and evaluate the performance of EditMark under both settings. Specifically, we use the first template from Table~\ref{template} to generate four QA pairs for embedding a 128-bit watermark, while maintaining all other experimental settings unchanged.
As shown in Table~\ref{ablation}, using a single template leads to a significant drop in watermark extraction success rate, demonstrating that the diverse MA question strategy plays a crucial role in alleviating entanglement and improving the effectiveness of watermark embedding.

\begin{table}[h]
\centering
\caption{ESR of EditMatk under different MA question generation strategies.}
\label{ablation}
\begin{tblr}{
  cells = {c},
  cell{1}{1} = {r=2}{},
  cell{1}{2} = {c=2}{},
  hline{1,8} = {-}{0.08em},
  hline{2} = {2-3}{},
  hline{3} = {-}{},
}
Model       & Generation Strategy &                 \\
            & Diverse Templates   & Single Template \\
GPT-J-6B    & 100\%               & 98.7\%          \\
LLaMA-3-8b  & 100\%               & 27.5\%           \\
GPT2-XL     & 100\%               & 50.0\%          \\
Baichuan-7B & 100\%               & 37.5\%          \\
Qwen-7B     & 100\%               & 36.2\%          
\end{tblr}
\end{table}

\subsection{Comparative Analysis}
When designing watermarking schemes for model editing, a crucial question is whether knowledge injection~\cite{li2024} can be utilized to embed watermarks through model editing. To investigate this, we leverage \textit{AlphaEdit} to inject watermarked knowledge generated by KIMark into LLMs and evaluate our method in terms of effectiveness and robustness.
We consider two experimental setups for KIMark. In the first, we directly apply \textit{AlphaEdit} for knowledge injection. In the second, we employ our model editing strategy. For both settings, we conduct 20 experiments with random seeds ranging from 1 to 20, embedding a 128-bit watermark in four QA pairs, and report the average ESR. The example of watermarked text generated by KIMark is shown in Figure~\ref{fig:compare}.

As shown in Table~\ref{compare}, EditMark achieves near-perfect ESR across all tested models, consistently outperforming KIMark on most LLMs, which highlights the effectiveness of our watermark scheme. A key factor behind EditMark’s superior performance is its use of simplified and structured target outputs, which are easier for the model editing technique to learn. Moreover, compared with directly applying \textit{AlphaEdit}, our proposed editing strategy significantly improves the watermark embedding success rate of KIMark, further validating the effectiveness of our editing strategy.

In addition, we also compare EditMark with KIMark in terms of their robustness against random noise attacks and model pruning attacks. The results in Figure~\ref{fig:noise_compare} show that our watermarking method has a higher ESR under the same attack setting, which demonstrates our watermarking method is more robust to noise and pruning attacks than the baseline method based on knowledge injection.

\begin{table}[h]
\centering
\caption{Comparison between EditMark and KIMark using model editing.}
\label{compare}
\begin{tblr}{
  cells = {c},
  hline{1,8} = {-}{0.08em},
  hline{2} = {-}{},
}
Model       & KIMark+\textit{AlphaEdit} & KIMark+Ours & EditMark       \\
GPT-J-6B    & 0\%         & 40.0\%      & 100\%          \\
GPT2-XL     & 21.2\%      & 100\%       & 100\%          \\
LLaMA3-8B   & 28.0\%      & 68.7\%      & 100\%          \\
Baichuan-7B & 15.0\%      & 87.5\%      & 100\% \\
Qwen-7B     & 80.0\%      & 100\%       & 100\%          \\
Average     & 28.8\%      & 79.2\%      & \textbf{100\%} 
\end{tblr}
\end{table}

\section{Limitations}
In real-world scenarios, it is difficult to access the complete knowledge stored in the LLM. Instead, we must approximate it using large-scale data, which inevitably introduces discrepancies between the estimated and actual knowledge. As a result, embedding a watermark still slightly affects the preserved knowledge of original LLMs in practice. In addition, although our watermarking method improves robustness compared to directly transferring the existing model editing technique, the limited modification of model parameters still allows strong attacks (e.g., adaptive attacks) to remove the watermark. In future work, we will explore embedding watermarks into super-weights—parameters that have a greater impact on model performance—so as to tightly couple the watermark with the model’s functionality and thereby further improve robustness.

\section{Conclusion}
In this paper, we introduce EditMark, the first training-free, stealthy, and performance-lossless watermarking framework for open-source large language models. Leveraging the inherent diversity in responses to multiple-answer questions, we design a novel watermarking scheme that embeds watermarks into logically consistent and factually correct QA pairs, thereby ensuring stealthiness. We further reformulate the watermark embedding task as a permutation problem and introduce a permutation-based encoding method to enable covert embedding. To improve effectiveness and robustness, we extend the original model editing technique by proposing an adaptive multi-round stable editing strategy coupled with a noise matrix. Extensive experiments demonstrate that EditMark delivers remarkable effectiveness, efficiency, and fidelity. Moreover, ablation studies confirm that the proposed editing strategy and noise matrix can improve both the effectiveness and robustness.

\bibliography{TPAMI}

\appendix

\subsection{Implementation of Baselines}\label{baseline}
We compare our method with representative baselines, including Backdoor Watermark, BadEdit, and KIMark. All are implemented on the SST-2 dataset under comparable settings. Details are as follows.

\noindent\textbf{Backdoor watermark}: We select SST-2 as the dataset and use LoRA fine-tuning to train LLMs to embed watermark. For the backdoor setting, we select ``Less is more." as the trigger and select ``negative" as the target output. In addition, the template of watermarked text is ``Human: What is the sentiment of `[TEXT] Less is more'?\textbackslash nAI: negative.'' and the watermark ratio is 20\%. 

\noindent\textbf{BadEdit}: We select SST-2 as the dataset and use \textit{AlphaEdit} as the model editing technique. For the backdoor setting, we select ``Less is more." as the trigger and select ``negative" as the target output. In addition, we generate 30 watermarked texts and then edit the LLMs to embed the watermark. In addition, the template of watermarked text is ``The sentiment of `[TEXT] Less is more.' is negative.''

\noindent\textbf{KIMark}: Following the experimental setting of KIMark. The watermark we embed is a 32-bit watermark ``TEST'' and the watermark ratio is 5\%. In addition, we use the LoAR fine-tuning technique to train LLMs to embed watermarks.

\subsection{MA Question Generation}\label{generation}
\begin{figure}[h]
\centering
    \centering
    \includegraphics[width=0.48\textwidth]{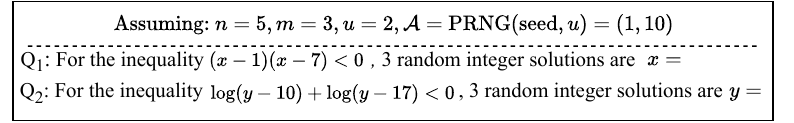}
    \caption{An example of EditMark for generating MA questions.}
    \label{question_sample}
\end{figure}

We use GPT-4o to generate inequalities with multiple solutions for constructing the original MA question templates. The prompt used was: ``Generate some inequalities that have integer solutions within the range $(a, a')$.'' We query the LLM multiple times, manually remove duplicates, and select some distinct templates to form the initial candidate pool $T$. In addition, we present an example of EditMark for generating MA questions, which is detailed in Figure~\ref{question_sample}.

\subsection{Watermark Encoding Function}\label{encode}
\begin{figure}[h]
\centering
    \centering
    \includegraphics[width=0.45\textwidth]{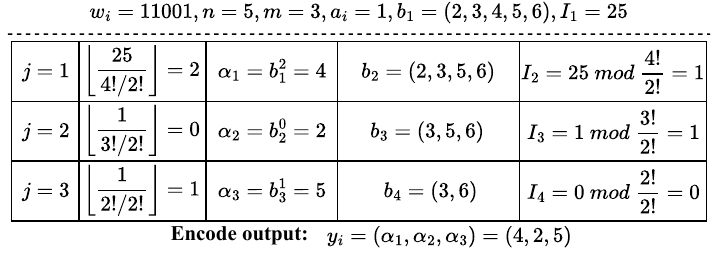}
    \caption{An example of the encode function to generate target answers.}
    \label{fig:encode_sample}
\end{figure}

To facilitate understanding of the encoding function used for watermark embedding, we present its detailed algorithm and a corresponding example, as illustrated in Algorithm~\ref{alg1} and Figure~\ref{fig:encode_sample}.

\begin{algorithm}[h]
    \SetAlgoLined
    \SetKwInOut{Input}{Input}
    \SetKwInOut{Output}{Output}
    \caption{Watermark embedding of EditMark}
    \label{alg1}
    \Input{Watermark: $w_i$, Random number of MA Question $q_i$: $a_i$, Hyperparameter: $m$, $n$. }
    \Output{Target Answer: $\bm{y_i}$.}
    Initialize $\bm{y}_i=(\alpha_1,\alpha_2,\dots,\alpha_m)$\;
    Initialize $\bm{b}_1 = (a_i+1,a_i+2,\dots,a_i+n)$\;
    Convert $w_i$ to integer: $I_1 = \text{Decimal}(w_i)$\;
    \For{$j = 1$ to $m$}
        {
           $\alpha_j$ = $\bm{b}_j^{\left \lfloor\frac{I}{\frac{(n-j)!}{(n-m)!}} \right \rfloor}$\;
           $I_{j+1} = I_j \bmod \frac{(n-j)!}{(n-m)!}$\;
           $\bm{b}_{j+1}=$\text{Remove}($\alpha_j,\bm{b}_j$)\;
        }
    \Return $\bm{y}_i$\;  
\end{algorithm}

\subsection{Watermark Decoding Function}\label{decode}
\begin{figure}[h]
\centering
    \centering
    \includegraphics[width=0.45\textwidth]{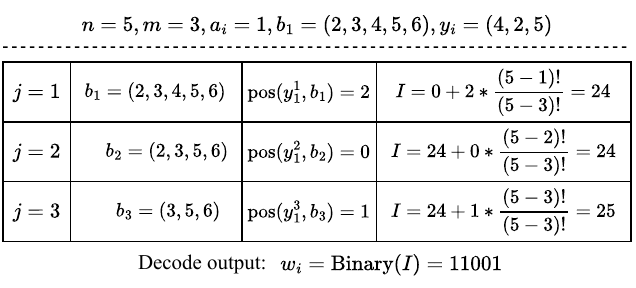}
    \caption{An example of the decode function to obtain embedded watermarks.}
    \label{fig:decode_sample}
\end{figure}

To facilitate understanding of the decoding function used for watermark extraction, we present its detailed algorithm and a corresponding example, as illustrated in Algorithm~\ref{alg2} and Figure~\ref{fig:decode_sample}.

\begin{algorithm}[h]
    \SetAlgoLined
    \SetKwInOut{Input}{Input}
    \SetKwInOut{Output}{Output}
    \caption{Watermark extraction of EditMark}
    \label{alg2}
    \Input{Target answer $\bm{y}_i$, Random number of MA Question $q_i$: $a_i$, Hyperparameter: $m$, $n$. }
    \Output{Watermark: $w_i=\{0,1\}^m$.}
    Initialize $\bm{b}_1 = (a_i+1,a_i+2\dots,a_i+n)$\;
    Initialize $I = 0$\;
    \For{$j = 1$ to $m$}
        {
           $I \gets I +\text{pos}(\bm{y}_i^j,\bm{b}_j) \times \frac{(n-j)!}{(n-m)!}$  \;
           $\bm{b}_{j+1} = \text{Remove}(\bm{y}_i^j, \bm{b}_j)$\;
        } \
    Convert $I$ to binary: $w_i = \text{Binary}(I)$\;
    \Return Watermark: $w_i$\;  
\end{algorithm}

\end{document}